\documentclass[12pt]{article}
\usepackage{amsmath}
\usepackage{graphicx}
\usepackage{enumerate}
\usepackage{natbib}
\usepackage{url} 
\usepackage{framed}

\usepackage{tikz}
\usepackage{amsmath}
\usetikzlibrary{arrows.meta}

\usepackage{graphicx,verbatim,array,multicol, palatino}
\usepackage{amsthm, amsmath, amssymb,  setspace,natbib}
\usepackage{epsfig, url,epstopdf}
\usepackage{indentfirst,enumerate}
\usepackage{color}
\usepackage[titletoc,title]{appendix}
\usepackage{booktabs}
\usepackage{graphicx}
\usepackage{placeins}
\usepackage{subcaption}
\usepackage{adjustbox}
\usepackage{tabularx, booktabs}

\definecolor{red}{rgb}{1,0,0}
\definecolor{blue}{rgb}{0,0,1}
\definecolor{green}{rgb}{0,0.6,0.4}

\def \boldeta{\boldsymbol{\eta}}

\usepackage{hyperref}
 \hypersetup{hypertex=true,
            colorlinks=true,
            linkcolor=blue,
            anchorcolor=red,
            menucolor=red,
            citecolor=blue}
\usepackage{tikz}
\usetikzlibrary{shapes,graphs}
\usetikzlibrary{automata,positioning}
\usetikzlibrary{arrows,	petri,topaths}
\usepackage{pgf}
\usepackage{tikz, pgfplots}
\usepackage[titletoc,title]{appendix}
\usepackage{algorithm,algpseudocode}
\usetikzlibrary{shapes,decorations,arrows,calc,arrows.meta,fit,positioning}
\tikzset{
    -Latex,auto,node distance =1 cm and 1 cm,semithick,
    state/.style ={ellipse, draw, minimum width = 0.7 cm},
    point/.style = {circle, draw, inner sep=0.04cm,fill,node contents={}},
    bidirected/.style={Latex-Latex,dashed},
    el/.style = {inner sep=2pt, align=left, sloped}
}

\newtheorem{theorem}{Theorem}

\newtheorem{lemma}{Lemma}
\newtheorem{assumption}{Assumption}
\newtheorem{example}{Example}

\newtheorem{Remark}{Remark}

\begin{document}

\title{\baselineskip=20pt
Quasi Instrumental Variable Methods for Stable\\ Hidden Confounding and Binary Outcomes}

\author{ Zhonghua Liu, Baoluo Sun, Ting Ye,\\ David Richardson,  Eric Tchetgen Tchetgen\thanks{Zhonghua Liu is Assistant Professor in the Department of Biostatistics at Columbia University ({\em Email: zl2509@cumc.columbia.edu}). Baoluo Sun is Assistant Professor  in the Department of Statistics and Data Science, National University of Singapore ({\em stasb@nus.edu.sg}). Ting Ye is 
Assistant Professor in the Department of Biostatistics, University of Washington, Seattle ({\em Email: 
tingye1@uw.edu}). David Richardson, Professor of Environmental and Occupational Health, Associate Dean of Research, University of California, Irvine ({\em Email:david.richardson@uci.edu}).  Eric Tchetgen Tchetgen is University Professor,
Professor of Biostatistics in Biostatistics and Epidemiology, 
Professor of Statistics and Data Science at  University  of Pennsylvania ({\em Email: ett@wharton.upenn.edu}).  
}}
\date{}
\date{}

\maketitle
\thispagestyle{empty}

\begin{abstract} 
\linespread{1}\selectfont
{

Instrumental variable (IV) methods are central to causal inference from observational data, particularly when a randomized experiment is not feasible. However, of the three conventional core IV identification conditions, only one, IV relevance, is empirically verifiable; often one or both of the other conditions, exclusion restriction and IV independence from unmeasured confounders, are unmet in real-world applications. These challenges are compounded when the outcome is binary, a setting for which robust IV methods remain underdeveloped. A fundamental contribution of this paper is the development of a general identification strategy justified under a structural equilibrium dynamic generative model of so-called \emph{stable confounding} and a \emph{quasi instrumental variable} (QIV), i.e. a variable that is only assumed to be predictive of the outcome. Such a model implies (a) stability of confounding on the multiplicative scale, and (b) stability of the additive average treatment effect among the treated (ATT), across levels of that QIV. The former is all that is necessary to ensure a valid test of the causal null hypothesis; together those two conditions establish nonparametric identification and estimation of the conditional and marginal ATT. To address the statistical challenges posed by the need for boundedness in binary outcomes, we introduce a generalized odds product re-parametrization of the observed data distribution, and we develop both a principled maximum likelihood estimator and a triply robust semiparametric locally efficient estimator, which we evaluate through simulations and an empirical application to the UK Biobank.
}

\vspace{.1in}
\end{abstract}

\newpage
\pagestyle{plain}
\setcounter{page}{1}

\section{Introduction}
\label{sec:introduction}

\begin{figure}[H]
	\begin{subfigure}[b]{.5\linewidth}
		\centering
		\begin{tikzpicture}
    \node[state] (1) {$Z$};
    \node[state] (2) [right =of 1] {$A$};
    \node[state] (3) [right =of 2] {$Y$};
    \node[state] (5) [above =of 2,xshift=.9cm, yshift=-0.2cm] {$U$};
    \path (1) edge node[above] {} (2);
    \path (2) edge node[above] {} (3);
    \path (5) edge node[el,above] {} (2);
    \path (5) edge node[el,above] {} (3);
  
\end{tikzpicture}
		\caption{}
		\label{subfig:DAG_valid_IV}
	\end{subfigure}%
	\begin{subfigure}[b]{.5\linewidth}
	\centering
	\begin{tikzpicture}
      \node[state] (1) {$Z$};
    \node[state] (2) [right =of 1] {$A$};
    \node[state] (3) [right =of 2] {$Y$};
    \node[state] (5) [above =of 2,xshift=.9cm, yshift=-0.2cm] {$U$};
    \path[draw=black] (1) edge node[above] {} (2);
    \path (2) edge node[above] {} (3);
    \path (5) edge node[el,above] {} (2);
    \path (5) edge node[el,above] {} (3);
    \path[draw=red,dashed] (5) edge node[el,above] {} (1);
    \path[draw =blue,densely dashed] (1) edge[bend right=40] node[el,above] {} (3);
    \end{tikzpicture}
		\caption{}
		\label{subfig:DAG_invalidIV}
	\end{subfigure}
	\caption[DAG]{Directed acyclic graphs depicting: (a) a valid IV; (b) an invalid IV in which case the IV independence and exclusion restriction assumptions do not 
		necessarily hold. Here, $Z$ is the candidate IV, $A$ is the exposure, $Y$ is the binary outcome of interest and $U$ represents unmeasured confounding factor.  The blue dashed line from $Z$ to $Y$ indicates the violation of the exclusion restriction assumption.  The red dashed line from $U$ to $Z$ indicates the violation of the IV independence assumption. }
	\label{fig:DAG_IV}
\end{figure}
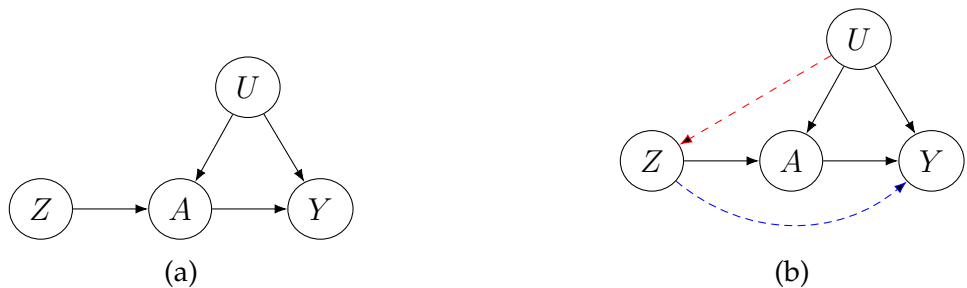

In observational studies, identification and estimation of causal effects such as the average treatment effect on the treated (ATT) can be particularly challenging because of concerns about confounding bias. 
As one can seldom guarantee that all relevant confounding factors have been measured in a given study, concerns about unmeasured confounding bias have led to several recent developments in causal inference. One such method, the instrumental variable (IV) approach, offers a viable path towards identification of ATT in presence of unmeasured confounding, under key identification conditions; for a recent example, see \citet{Liu2020Sinica}. Specifically, a valid IV must satisfy three conventional core assumptions: (IV.1) it must be associated with the treatment variable (IV relevance);  (IV.2) it must affect the outcome only through the treatment (exclusion restriction); and (IV.3) it must be independent of unmeasured confounders of the treatment-outcome relationship (IV independence). The causal directed acyclic graph (DAG) in Figure (\ref{subfig:DAG_valid_IV})  provides a graphical depiction of these three core IV assumptions. 

While assumptions (IV.1) -- (IV.3) can be used to construct a valid test of the sharp null hypothesis of no causal effect, or to obtain valid bounds for the ATT, they fail to provide point identification of the ATT without an additional condition.  \citet[Chapter 16]{Hernan-Robins2020whatif} provide a detailed discussion on the possible nature of such a fourth condition to achieve point identification; while \citet{Liu2020Sinica} characterize a necessary and sufficient condition which together with (IV.1)--(IV.3) identifies the distribution of the treatment-free potential outcome among the treated, and therefore the ATT for any outcome type.   A prominent example of such a fourth condition in the literature which happens to be central to this paper is an additive homogeneity condition which states that the ATT is constant across levels of the IV, also known as \emph{no-current treatment value interaction assumption} \citep{Robins1989,Robins1994,Hernan2006,Hernan-Robins2020whatif}. This condition essentially rules out any additive interaction between an unmeasured confounder and the treatment under a structural model for the outcome among the treated \citep{Hernan2006}. \citet{Liu2020Sinica} proposed an alternative identification strategy by restricting the degree of heterogeneity in a so-called extended propensity score function which encodes residual confounding by explicitly spelling out the dependence of the treatment mechanism on the treatment-free potential outcome. Also see \citet{liu2025multiplicative} for recent developments for identification and inference about the ATT under a so-called multiplicative IV model.

Several instrumental variable (IV) methods for causal inference with binary outcomes have been proposed in the literature \citep{angrist2001estimation,Robins2004,vansteelandt2011instrumental,clarke2012instrumental,swanson2018partial}. These methods may be biased if any of core IV assumptions (IV.1)--(IV.3) does not hold. The potential for violation of any of these conditions is a valid concern in practice; for example, so-called Mendelian randomization (MR) studies, a popular form of IV analysis in epidemiology, in which genetic variants (e.g., single-nucleotide polymorphisms or SNPs) are used as instrumental variables, is susceptible in practice to violations of core IV assumptions \citep{Katan1986,Smith2003a}.

  Of the three core IV assumptions (IV.1)-(IV.3),  only IV relevance (IV.1) is empirically verifiable by assessing the association between the IV and the exposure of interest. For instance, in MR settings this is typically done by confirming such associations from the catalog of genome-wide association studies (GWAS) at \url{https://www.ebi.ac.uk/gwas/}.  Unfortunately, assumptions (IV.2) and (IV.3) are not empirically verifiable and can only be falsified  in special settings \citep{Bonet2001, Glymour:2012aa, swanson2018partial}, although such tests can be under-powered. While in other settings, for example when the exposure variable is continuous,  such empirical falsification tests may not exist \citep{Bonet2001}.

A limited but growing number of methods have been developed for valid inference using an IV despite possible violations of assumptions (IV.2) and/or (IV.3)  for a continuous outcome in the single IV setting
 \citep{Spiller2019detecting, Spiller2020,tchetgen2021genius,sun2022selective,ye2024genius,liu2022mendelian,kang2024identification}. 
  These robust MR methods apply primarily to continuous outcomes, however they do not fully address the unique methodological challenges associated with binary outcomes. First, it is not immediately clear that these methods can uniquely identify the ATT in analyses of binary outcomes under violations of (IV.2) and/or (IV.3) given that they fundamentally rely on a structural linear model for the outcome. Second, modeling challenges are further accentuated by the presence of nuisance functions which can be difficult to model in a coherent fashion such that they stand a chance for consistent estimation, while ensuring that the resulting ATT estimator lies in its natural parameter space $(-1,1)$ for a binary outcome. Third, concerns about potential model misspecification bias warrants the development of new robust and efficient semiparametric estimators for the causal effect of interest.

  All three challenges are addressed in this paper. 
Specifically, a main contribution of this work is the development of a novel \emph{Quasi-Instrumental Variable} (QIV) identification strategy justified under a structural equilibrium dynamic generative model with hidden confounding, by leveraging any QIV that is only required to be predictive of the outcome among the untreated. Such a model implies (a) stability of confounding bias on the multiplicative scale, as well as (b) stability of the additive treatment effect across levels of the QIV. Under this framework, we propose a new test of the causal null hypothesis of no conditional ATT under (a); furthermore, we establish nonparametric identification of the conditional and marginal ATT under both (a) and (b). The fact that their conjunction conveys robustness against violation of standard IV conditions appears to be new. Another key contribution of the paper is the description of a structural equations  equilibrium dynamic generative model which justifies (a) and (b), provided that the realized data sample can reasonably be viewed as obtained from the equilibrium distribution of a certain hidden Markov process; see Section \ref{sec:edg} for a detailed discussion.

Beyond identification, 
inference under (a) and (b) presents new significant modeling challenges, as the ATT for a binary outcome defines a causal risk difference and therefore must fall in $(-1,1)$, while an observed data model compatible with (a) must also be compatible with the counterfactual risk under no treatment for both the treated and untreated in $(0,1)$.  These natural restrictions imply that the range of possible values for the baseline risk necessarily depends on the range of possible values for the ATT and a function encoding the degree of  confounding bias (a numerical example is given in Section \ref{sec:estimation} to illustrate this difficulty). Consequently, the joint range of the baseline risk, the ATT and the function encoding confounding bias is strictly smaller than the Cartesian product of their marginal ranges. In other words, the parameters in the baseline risk model, the ATT and the confounding bias are variation dependent on each other, introducing an extra-layer of difficulty for parameter estimation.  To resolve this issue,  we propose to generalize the odds product model formulation introduced by \citet{Richardson2017} to re-parameterize the nuisance models such that model parameters are variation independent, which opens up an opportunity to perform formal estimation techniques ranging from unconstrained parametric maximum likelihood estimation to more sophisticated semiparametric estimators, while ensuring that the resulting estimator of the ATT is always contained in $(-1,1)$.

 To further increase robustness to possible model misspecification bias, we propose a {\it triply robust} semiparametric estimator that is consistent and asymptotically normal (CAN) provided some, but not necessarily all posited nuisance models are correctly specified. We then show that the proposed triply robust estimator attains the semiparametric local efficiency bound when all working models are correctly specified.  
Finally, we evaluate the finite sample performance of the proposed methods in extensive simulation studies and in an application evaluating the causal effect of adiposity on the risk of hypertension in the UK Biobank.

The rest of the paper is organized as follows. In Section \ref{sec:identify}, we introduce notation and the basic generative equilibrium structural equation model which motivates our identifying conditions. In Section \ref{sec:estimation}, we describe maximum likelihood and triply robust semiparametric estimators using our proposed generalized odds product parametrization. The simulation exercise is reported in Section \ref{sec:simulation}, while Section \ref{sec:ukb} provides an illustration of the methods in the UK Biobank. The paper ends with a discussion in Section \ref{sec:discussion}.

\section{QIV-based Test and Identification}
\label{sec:identify}
\subsection{Problem Setup}
{Suppose that one has observed data $O=(Y,A,Z,{X})$, where $Y$ is a binary outcome of interest, $A$ a binary exposure/treatment variable, $Z$ a pre-exposure, possibly invalid binary instrument, and $q$-dimensional baseline covariates ${X}=(X_1,...,X_q)$.  Throughout, $U$ will denote a set of unmeasured confounders such that, conditioning on $X$ and $U$ would in principle suffice to identify the causal effect of $A$ on $Y$.  Let $P(O)$ denote the density of the observed data distribution with respect to the product of the counting measure on $\{0,1\}^3$ and $\mu=\otimes_{j=1}^q \mu_j$, where $\mu_j$ is some appropriate dominating measure for the marginal distribution of $X_j$.} 
Let $Y_a$  denote the potential outcome had possibly contrary to fact, $A$  been set to $a \in \{0,1\}$ and let $Y_{az}$ denote the potential outcome had  $A$ and  $Z$  been set to $a$ and $z$ respectively, $(a,z)^T \in \{0,1\}^2$. The causal estimand of interest is the following marginal average  treatment effect on the treated (ATT) defined on the additive scale 
\begin{equation*}
\gamma= E(Y_{a=1}- Y_{a=0}\mid A=1), \gamma \in (-1,1).
\end{equation*}

We make the  consistency assumption: $Y = AY_{a=1} + (1-A)Y_{a=0}$, which links the observed outcome to the potential outcome. 
Under  consistency,  $E(Y_{a=1}| A=1)$ is equal to  $E(Y|A=1)$, which can thus be consistently estimated by the average observed outcome among treated individuals. The observed  risk difference between treated and untreated groups is related to the ATT and the confounding bias within strata of $X=x$ by:

 \begin{align*}
\underbrace{E(Y| A=1,{X}={x}) - E(Y| A=0,{X}={x})}_{\text{Observed risk difference}} =& \underbrace{E(Y_{a=1} - Y_{a=0}| A=1,{X}={x})}_{\text{Conditional ATT}}\\ &+ \underbrace{E(Y_{a=0} | A=1,{X}={x}) - E(Y_{a=0} |A=0,{X}={x})}_{\text{Confounding bias}}.
\label{eqn:obs-ATT-selection}
 \end{align*}
 The conditional expected value of the  potential outcome $Y_{a=0}$ among  treated individuals $E(Y_{a=0}\mid A=1,{X}={x})$   is not identified because A is potentially subject to unmeasured/hidden confounding even after conditioning on ${X}$; this phenomenon is sometimes referred to as the selection problem in the econometrics literature \citep{angrist2008mostly}, and is also reflected by  the confounding bias term
being nonzero.
 In other words, the treatment variable $A$ and the treatment-free potential outcome $Y_{a=0}$  are associated conditional on ${X}$, reflecting the presence of confounding bias due to the presence of $U$, a hidden common cause of the treatment and outcome.

\subsection{Initial Identification Strategy with an Imperfect IV }

It is clear that ATT cannot be identified with observed data $(A,Y,{X})$  in the presence of unmeasured confounding without an additional condition.  For $z\in \{0,1\}$, let  $\delta(z,x) = E(Y_{a=0}|A=1,Z=z,X=x) - E(Y_{a=0}|A=0,Z=z,X=x)$  denote the confounding bias on the additive scale, and the conditional ATT is  $\gamma \left( z,x\right) = E( Y_{a=1}|A=1,Z=z,X=x) - E(
	Y_{a=0}|A=1,Z= z,X=x) \in \left( -1,1\right)$.   The observed risk difference is simply equal to  the sum of the  conditional ATT $\gamma(z,x)$ and conditional confounding bias  $\delta(z,x)$ as follows:
\begin{eqnarray}
    E(Y| A=1,Z=z,X=x) - E(Y| A=0,Z=z,X=x) = \gamma(z,x) + \delta(z,x).
    \nonumber
\end{eqnarray}
\citet{robins1999sensitivity} proposed the following re-parameterization of the outcome conditional mean function in terms of the $\gamma(z,x)$ and $\delta(z,x)$:
\begin{equation*}
    E(Y|A=a,Z=z,X=x) = \gamma(z,x)a + \delta(z,x)\{a-P(A=1|z,x)\} + E(Y_{0z}|z,x).
\end{equation*}
In addition to the three core IV assumptions (IV.1)--(IV.3) defining a valid IV, which imply $E(Y_{a=0,z}|z,x)=E(Y_{a=0}|z,x)=E(Y_{a=0}|x)$, \citet{Robins1994}  makes  the following \textit{no current treatment value interaction }assumption to identify  ATT: $\gamma(z,x) = \gamma(x)$, that is, the ATT is stable across values of $z$ and therefore does not vary with $z$. \citet{tchetgen2013alternative} instead assume that the confounding bias  function $\delta(z,x)$ does not vary with $z$ in order to identify $\delta(x)$ and an unrestricted conditional ATT $\gamma(x,z)$. \citet{Liu2020Sinica} proposed an alternative identification strategy based on a stability condition of an odds ratio analog of the confounding bias function. 

However, the aforementioned identification strategies rely on having in hand a valid IV \( Z \) that satisfies assumptions (IV.1)--(IV.3). As we have argued, in practice \( Z \) may often fail to meet either the IV independence or exclusion restriction conditions, rendering previous identification strategies unable to identify the ATT.
It is also instructive to briefly illustrate identification obtained via so-called difference-in-differences (DiD) methods using the above parametrization. Thus, consider a standard repeated cross-sectional DiD set up where \(Z = T\) is a
binary variable indexing whether a unit's outcome is observed at a time \(T = 0\) or at time \(T = \)1.  In a standard DiD study design, treatment is assigned at a specific time strictly between \(T=0\) and \(T=1\), in which case a standard no anticipation condition implies that $\gamma(z=0,x) =0$ since the treatment cannot causally impact an outcome observed before it is assigned. Furthermore, the standard parallel trend assumption typically assumed in DiD settings is equivalent to the no interaction assumption of \citet{tchetgen2013alternative}: $\delta(z,x)={\delta}(x)$. It is then straightforward to check that the conjunction of these two conditions provides identification of $\gamma(z=1,x) =0$ via the standard conditional DiD estimand $\sum_{a,z}(-1)^{a+z}E(Y|A=a,Z=z)$; therefore, DiD combines the no anticipation condition with parallel trends to identify a causal effect, without invoking the exclusion restriction (IV.2); that is DiD allows for $E(Y_{a=0}|z,x)$ to remain unrestricted, a desirable robustness property we wish for our QIV to fulfill. It is further worth noting that in repeated cross-sectional settings where it is not possible to ensure that the treatment assignment strictly follows the first outcome measurement, the no anticipation assumption $\gamma(z=0,x) =0$ generally would not hold and thus the standard DiD estimand may fail to identify the conditional ATT.

An initial strategy to relax (IV.2) and (IV.3) inspired by the above exposition, replaces both conditions with the alternative conditions of stable additive ATT and confounding bias functions with respect to $z$: $\gamma(z,x) = \gamma(x), \delta(z,x) = \delta(x)$; essentially combining the stability conditions of \citet{Robins1994} and \citet{tchetgen2013alternative}. This strategy clearly is also relevant to consider in the above DiD setting without imposing the no anticipation assumption. Either way, it is straightforward to show that $E(Y|A=a,Z=z,X=x)$ would then reduce to:
\begin{equation*}
    E(Y|A=a,Z=z,X=x) = \{\gamma(x)+ \delta(x)\}a
    +\omega(x,z).
\end{equation*}
where $\omega(x,z)=E(Y_{0z}|z,x)-P(A=a|z,x)\delta(x)$ so that it is not possible to tease $\gamma(x)$ apart from $\delta(x)$. To see this, note that the above equation is equivalently expressed as \begin{equation*}
    E(Y|A=a,Z=z,X=x) = \{\gamma^*(x)+ \delta^*(x)\}a
    +\omega^*(x,z).
\end{equation*}
where $\gamma^*(x)=\gamma(x)+v(x);\delta^*(x)=\delta(x)-v(x);\omega^*(x,z) =E(Y_{0z}|z,x)+P(A=1|z,x)v(x)+ P(A=1|z,x)\delta^*(x)$ for any $v(x)$, proving the lack of identification of $\gamma(x)$ under this strategy. Below, we propose a careful modification of this initial strategy which successfully leads to identification, by instead considering stability conditions that are operating on two different scales, potentially providing a principled solution to both settings of an imperfect instrument, or of DiD where no-anticipation fails.

\subsection{An Equilibrium Dynamic Generative Model of Stable Hidden Confounding}

\label{sec:edg}
\begin{figure}
    \centering

\begin{tikzpicture}[
  node distance=1cm and 1cm, 
  every node/.style={minimum size=0.5cm},  
  every path/.style={->, >=Stealth, thick}
  ]
  \node (A0) {$A(0)$};
  \node[right=of A0] (A1) {$A(1)$};
  \node[right=of A1] (dots) {$\dots$};
  \node[right=of dots] (At1) {$A(T-1)$};
  \node[right=of At1] (AT) {$A(T)$};
  \node[above=of A0] (U0) {$U(0)$};
  \node[above=of A1] (U1) {$U(1)$};
  \node[above=of At1] (UT1) {$U(T-1)$};
  \node[right=of AT] (Y) {$Y$};
  \node[below= 1.5cm of dots] (Z) {Z};  
  \node[above=of dots] (adots) {$\dots$}; 

  \draw (Z) -- (A0);
  \draw (Z) -- (A1);
  \draw (Z) -- (At1);
  \draw (Z) -- (AT);

  \draw (U0) -- (A1);
  \draw (A1) -- (U1);
  \draw  (At1) -- (UT1);
  \draw (UT1) -- (AT);
  \draw (UT1) -- (Y);
  \draw (U1) -- (dots);
\draw[dashed] (Z) -- (Y);
  \draw (AT) -- (Y);
\end{tikzpicture}
\caption{
Causal directed acyclic graph (DAG) illustrating the latent equilibrium dynamic model for treatment and unmeasured confounding.}
\label{figure:markovmodel}
\end{figure}
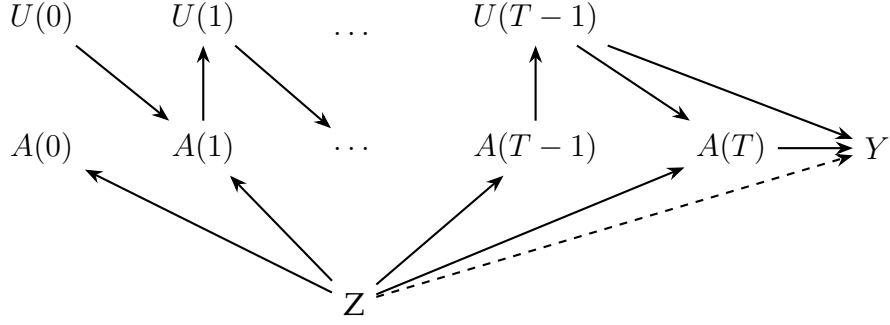

Our testing and identification framework is inspired by a general equilibrium dynamic generative (EDG) model, in which the 'quasi-instrument' $Z$ is realized at baseline ($t = 0$) and may affect the evolution of treatment $A(t)$ and unmeasured confounder $U(t)$ over time. The corresponding  causal directed acyclic graph (DAG) is depicted in Figure \ref{figure:markovmodel}. By time $T$, the system reaches a steady state, such that the joint distribution of $(A(T), U(T{-}1))$ is stable over time. The outcome $Y$ is determined by the most recent exposure and confounder. Under this equilibrium assumption, the distribution of $U$ becomes independent of $Z$ conditional on $A$, justifying our key assumptions of (a) stability of confounding on the multiplicative scale, and (b) stability of the additive treatment effect among the treated across levels of the QIV $Z$.

The EDG model best captures causal mechanisms in which behaviors and exposures might shift over time due to personal adaptation, environmental pressures, or policy influences only to eventually settle at an equilibrium at which such external and internal forces are at steady state. For example, we can assume that both an individual's exposure status ($A$) and unmeasured lifestyle-related confounders ($U$)—such as diet, physical activity, or smoking—arise from a dynamic process that changes gradually but whose distribution eventually stabilizes.

 The notion that a population distribution under which data are sampled is at an equilibrium is neither new nor particularly controversial, as it is a central tenet of several existing scientific theories, including the notion of Hardy-Weinberg Equilibrium in population genetics which states that the genetic variation in a population remains constant from one generation to the next in the absence of major disturbing factors \citep{edwards2008gh}; as well as Malthusian Equilibrium Theory which describes a population that is in balance with its resources \citep{malthus1798,galor2000}. This example suggests that our key identification assumptions are  reasonable in settings where the population dynamics generating the observed data can reasonably be assumed to have reached a \emph{Steady State}, in the sense that the underlying data generating law for the treatment assignment mechanism and confounding factors is stable over the time frame under consideration. 
 
 For example, in the application we later consider in Section \ref{sec:ukb}, our model essentially assumes that a person's hypertension status is observed at an equilibrium state (at least locally) within a specific time-frame of their lifecourse. The assumption would not be satisfied, in a population in which, the data are observed shortly after a highly effective intervention to reduce the prevalence of hypertension is introduced disrupting hypertension prevalence. However, the assumption may still be reasonable after some time, once the population has reached a new state of equilibrium with respect to hypertension prevalence. Finally, we note that an alternative interesting interpretation of the Markov conditions implied by the assumed DAG at equilibrium is that of the infinite DAG interpretation of a so-called Chain Graph model described in \citet{lauritzen2002chain}.

To formalize this idea, we now describe a structural equations model. For simplicity, we suppress baseline covariates $X$ and let $p_{aa^{\ast}}(z) = P(Y_a = 1 \mid A = a^{\ast}, Z = z)$. Consider the following outcome structural model that includes the unmeasured confounder of the $A$--$Y$ association, $U$,:

\begin{eqnarray*}
P \left( Y=1|A=a,U=u,Z=z\right)  =\beta _{a}\left( u\right) a+\beta
_{u,z}\left( u,z\right) 
,
\end{eqnarray*}%
where 
\[
A\amalg Y_{a}\mid U,Z,
\]
such that $(U,Z)$ is sufficient to account for confounding. While the model clearly accommodates exclusion restriction violations by allowing $\beta
_{u,z}\left( U,Z\right)-\beta
_{u,z}\left( U,0\right) \neq 0$, it appears to rule out any additive interaction between  $A$ and $Z$ in causing $Y$ conditional on $U$, that is $E(Y_{a=1,z,u}-Y_{a=0,z,u}\mid U=u,Z=z)$ does not depend on $z$. However, we do not view this as a real limitation of the model given that $U$ is not required to be observed and may reasonably be assumed to be enriched with all relevant effect modifiers of the additive effect of $A$ on $Y$, $\beta _{a}\left( u\right)\in (-1,1)$, which is otherwise {\it a priori} unrestricted. Furthermore, the no-additive interaction is guaranteed to hold under the sharp null hypothesis of no treatment effect. We also note that in the event $Z$ is a valid instrument, the structural regression model $P \left( Y=1|A=a,U=u\right)  =\beta _{a}\left( u\right) a+\beta
_{u}\left( u\right)$ is in fact completely unrestricted, other than satisfying the natural constraint that the left hand-side falls within the unit interval $(0,1)$ for all values of $(u,a)$. 

 In what follows, we will further presume that the model rules out any multiplicative interaction between $Z$ and $U$ in causing $Y_{a=0}$:
\[
P \left( Y_{a=0}=1|U=u,Z=z\right)/P \left( Y_{a=0}=1|U=u,Z=0\right) =\beta _{u,z}\left( u,z\right)=
\beta_{z}\left( z\right) 
\]
does not depend on $u$, where $\beta _{z}\left( 0\right) =1$. That is, $\beta _{u,z}\left( u,z\right)=\beta _{u}\left( u\right)\beta _{z}\left(z\right)$.
Interestingly, as shown in Section S2 of the Supplementary Materials (“Conditional independence justification for the no-interaction assumption”), this assumption can be motivated by a conditional independence condition involving counterfactual response types.

Finally, we presume that each i.i.d realization $(A,U)$ is a point snapshot of a latent dynamic Markov process, i.e. a unit's underlying full data actually entails a Markov process $\{A(t),U(t):t\}$  of time-varying treatment subject to time-varying confounding as represented in the DAG given in Figure \ref{figure:markovmodel}, where, without loss of generality $Z$ is assumed to be realized at $t=0$. The model then assumes that by the time $T$ when the treatment is measured, and thus $A\equiv A(T)\,\ $and $U\equiv U(T-1)$ the Markov process has reached a certain equilibrium state,  in the sense that its underlying generative law is stable over time, which can formally be stated as the following distributional equality:
\[
\left( A\left( T\right) ,U(T-1)\right) |Z\overset{d}{=}\left( A\left(
T-1\right) ,U(T-1)\right) |Z.
\]%
 Under these conditions,  
by d-separation
\[
Z\amalg U(T-1)|A(T-1).
\]%
We have that at equilibrium 
\[
Z\amalg U(T-1)|A(T),
\]%
and therefore 
\begin{eqnarray*}
&&P \left( Y=1|A=a,Z=z\right)  \\
&=&E\left\{P \left( Y=1|U,A=a,Z=z\right) |A=a,Z=z\right\}  \\
&=&E\left\{ \beta _{a}\left( U\right) |A=a,Z=z\right\} a+E\left\{ \beta
_{u}\left( U\right) |A=a,Z=z\right\} \beta _{z}\left( z\right)  \\
&=&E\left\{ \beta _{a}\left( U(T-1)\right) |A(T)=a,Z=z\right\} a+E\left\{ \beta
_{u}\left( U\left( T-1\right) \right) |A(T)=a,Z=z\right\} \beta _{z}\left(
z\right)  \\
&=&\gamma(z)a+\alpha^a(z)p_{00}\left( z\right),
\end{eqnarray*}%
where $p_{00}\left( z\right) =P \left( Y_{0}=1|A=0,Z=z\right)$, 
\begin{eqnarray*}
\alpha(z) &=&\frac{E\left\{ \beta _{u}\left( U\left( T-1\right) \right)
|A(T)=1,z\right\} }{E\left\{ \beta _{u}\left( U\left( T-1\right) \right)
|A(T)=0,z\right\} } \\
&=&\frac{E\left\{ \beta _{u}\left( U\right) |A=1,z\right\} }{E\left\{ \beta
_{u}\left( U\right) |A=0,z\right\} } \\
&=&\frac{P \left\{ Y_{a=0}=1|A=1,z\right\} }{P \left\{ Y_{a=0}=1|A=0,z%
\right\} } \\
&=&\frac{P \left\{ Y_{a=0}=1|A=1\right\} }{P\left\{ Y_{a=0}=1|A=0\right\} }\\
&=&\alpha,
\end{eqnarray*}%
and
\begin{eqnarray*}%
\gamma(z) &=&E\left\{ \beta _{a}\left( U(T-1)\right) |A(T)=1,Z=z\right\}  \\
&=&E\left\{ \beta _{a}\left( U\right) |A=1,Z=z\right\}  \\
&=&E\left\{ \beta _{a}\left( U\right) |A=1\right\}  \\
&=&E\left( Y_{a=1}-Y_{a=0}|A=1\right)\\
&=&\gamma;
\end{eqnarray*}%
such that $\frac{P \left\{ Y_{a=0}=1|A=1,z\right\} }{P \left\{ Y_{a=0}=1|A=0,z%
\right\} }$ and $E\left( Y_{a=1}-Y_{a=0}|A=1,z\right)$ do not depend on $z$. The first condition is a stability condition about the degree of multiplicative confounding as a function that is independent of the QIV, while the second is a stability condition about the degree of effect heterogeneity of the ATT on the additive scale. Both can therefore be seen as implied by the EDG model, although it is important to note that the converse is not necessarily true; i.e. both conditions do not imply the EDG model. As we discuss further in the next Section, the first stability condition can also be interpreted as a familiar nonlinear parallel trends condition with respect to $Z$ \citep{wooldridge2023simple}, in this case a multiplicative version of the parallel trends condition of \citep{tchetgen2013alternative} and standard DiD \citep{wooldridge2023simple};  while the second recovers 
the well-known {\it no current treatment value interaction} assumption of \citet{Robins1994}; both conditions are now directly implied by our Structural EDG model.    
We further expound upon these conditions below as a basis for credible inference about causal effects in presence of intractable but stable confounding.

\subsection{QIV-based Test of the Causal Null Hypothesis}

We propose a novel test of the causal null hypothesis,
$$H_0: \gamma(z,x)=0,\textup{ for all }z,x,$$
by leveraging a QIV. 
Our key idea is to encode the confounding bias function that captures the association  between $A$ and $Y_0$ conditional on $(Z,X)$  on the multiplicative scale:
\begin{equation*}
\alpha \left( z,x\right) =\frac{ E( Y_{a=0}|A=1,Z=z,X=x) }{E( Y_{a=0}|A=0,Z=z,X=x) }\in \mathbb{R}^{+},z=0,1.  \label{eq:alpha_z}
 \end{equation*}

Then we have
 \begin{equation}
 \label{eq:np}
 E(Y| A=1,Z=z,X=x) = \gamma(z,x)  + \alpha(z,x) E(Y| A=0,Z=z,X=x), 
 \end{equation}
which holds nonparametrically for $z=0,1$. We will assume that $Z$ serves as a predictive prognostic factor for the binary outcome $Y$ among the untreated \citep{hansen2008prognostic}.

\begin{assumption}[Relevance]
For each $x$, the quasi-instrumental variable $Z$ is predictive of the outcome among the untreated:
\[
E(Y \mid A=0, Z=1, X=x) - E(Y \mid A=0, Z=0, X=x) \neq 0.
\]
\label{assum3}
\end{assumption}
Assumption~\ref{assum3} is empirically testable and plays a role analogous to the classical IV relevance assumption (IV.1), as well as the outcome heteroscedasticity condition discussed in \citet{liu2022mendelian}. A violation—or near-violation—of this assumption may arise when $Z$ is only weakly associated with the outcome among untreated individuals, potentially resulting in weak identification bias \citep{liu2022mendelian,ye2024genius}. 

The QIV-based test of the causal null hypothesis  leverages the following multiplicative parallel trends assumption implied by the EDG model. 

\begin{assumption}[Multiplicative parallel trends]
The confounding bias on the multiplicative scale is stable, i.e. constant, across levels of the quasi-instrumental variable $Z$. That is, $\alpha(z=1,X)=\alpha(z=0,X)$, a.s., where 
\[
\alpha(z,x) = \frac{E(Y_0 \mid A=1, Z=z, X=x)}{E(Y_0 \mid A=0, Z=z, X=x)} = \alpha(x), \quad \text{for all } z \in \{0,1\}.
\]
\label{assum2}
\end{assumption}

Assumption \ref{assum2} states that conditional on $X$, the degree of multiplicative confounding bias remains constant across levels of $Z$ and therefore at most varies with $X$. Therefore, while the assumption relaxes the assumption of no hidden confounding bias, such bias is assumed to be balanced with respect to $Z$ as implied by  the  EDG model. 

\begin{figure}[h!]
\centering
\begin{tikzpicture}[>=Stealth, scale=1.1]
  \begin{scope}[xshift=2cm]

    \draw[->] (0,0) -- (6.5,0) node[right] {$Z$};
    \draw[->] (0,0) -- (0,4.5) node[above] {$\log E(Y_0 \mid A, Z, X=x)$};

    \node at (1.5,-0.3) {$Z=0$};
    \node at (4.5,-0.3) {$Z=1$};

    \draw[thick,red] (1.5,2.2) -- (4.5,3.2) node[right] {\small \textcolor{red}{$A=1$}};
    \draw[thick,blue] (1.5,1.2) -- (4.5,2.2) node[right] {\small \textcolor{blue}{$A=0$}};

    \filldraw[red] (1.5,2.2) circle (2pt);
    \filldraw[red] (4.5,3.2) circle (2pt);
    \filldraw[blue] (1.5,1.2) circle (2pt);
    \filldraw[blue] (4.5,2.2) circle (2pt);

    \draw [decorate,decoration={brace,amplitude=8pt}] (1.3,1.2) -- (1.3,2.2) node[midway,left=6pt] {\footnotesize $\log \alpha(x)$};
    \draw [decorate,decoration={brace,amplitude=8pt}] (4.3,2.2) -- (4.3,3.2);

  \end{scope}
\end{tikzpicture}
\caption{Multiplicative parallel trends: The log-risk of untreated potential outcomes for treated ($A=1$) and untreated ($A=0$) units follows parallel trends across $Z$. The constant vertical gap $\log \alpha(x)$ reflects proportional confounding, or multiplicative equi-confounding.}
\end{figure}
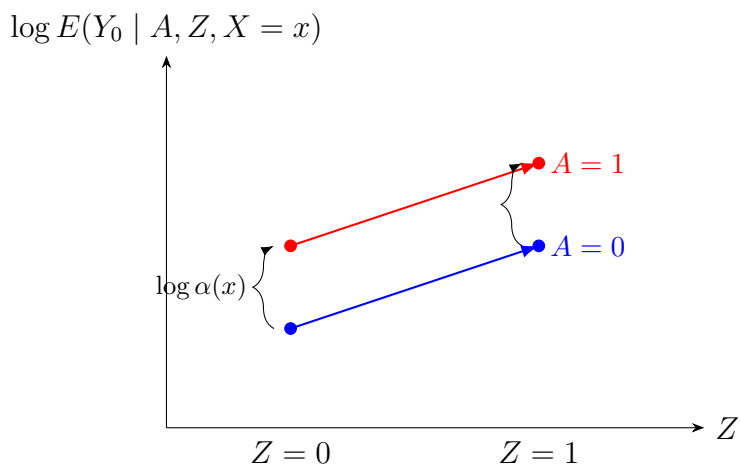

From equation (\ref{eq:np}), Assumptions~\ref{assum3} and~\ref{assum2} in conjunction with $\gamma(z,x)=0$ implies that the conditional distribution for the outcome $Y$ can be captured by
\begin{eqnarray*}
E(Y| A=a,Z=z,X=x) = \alpha(x)^a E(Y| A=0,Z=z,X=x),
	\end{eqnarray*}
for all $a$, $z$ and $x$, where
\begin{eqnarray*}
\alpha(x)=\frac{E(Y|A=1,Z=1, X=x)  -E(Y|A=1,Z=0,X=x)} {E(Y|A=0,Z=1,X=x)-E(Y|A=0,Z=0,X=x)}.
	\end{eqnarray*}
This may be used to construct a QIV-based score test of the causal null hypothesis $H_0$ from the observed data. When $X$ is multivariate with continuous components, we  outline a doubly robust score-type test  in Section \ref{sec:DR.est} which has nominal type 1 error even under partial misspecifications of the models for $E(Y| A=0,Z=z,X=x)$ or $P(A,Z\mid X=x)$.

\subsection{QIV-based Identification of Causal Effects}

For each fixed value of $x$, there are only two equations in (\ref{eq:np}) but three unknown parameters under Assumptions~\ref{assum3} and~\ref{assum2}: $\{\gamma(z,x),\alpha(x);z=0,1\}$. 
Therefore, an additional assumption is needed to reduce the number of unknown parameters to two, leading to point identification. Specifically, we consider {\it no current treatment value interaction} \citep{Robins1994} as implied by our EDG model.

\begin{assumption}[No current treatment value interaction]
The conditional ATT is invariant across levels of the QIV $Z$, that is,
\[
\gamma(z, x) = E(Y_1 - Y_0 \mid A=1, Z=z, X=x) = \gamma(x) \quad \text{for all } z.
\]
\label{assum1}
\end{assumption}

It is important to note that  Assumption \ref{assum1} allows for 
the presence of a  direct effect of $Z$ on $Y$ (the direct arrow from $Z$ to $Y$ is allowed to be present in Figure \ref{subfig:DAG_invalidIV}), i.e. $$E(Y_{a=0, z=1}- Y_{a=0, z=0}\mid A=1, Z=z,X=x)\neq 0,$$ and allows for the presence of  unmeasured confounding of the causal effects of $Z$ on $(A,Y)$, that is, violation of the IV independence assumption. 
Assumption \ref{assum1} is guaranteed to hold under the null hypothesis that the ATT is null for all treated individuals \citep{Hernan2006,wang2018bounded, liu2022mendelian}. We now summarize our main identification result in the following Theorem \ref{thm: nonpara-identi-multipli}.

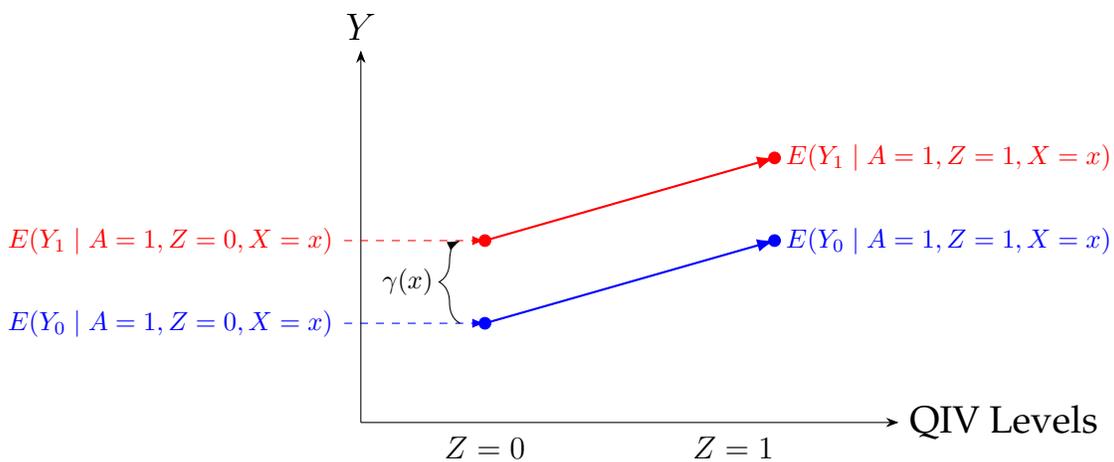
\begin{figure}[h!]
\centering
\begin{tikzpicture}[>=Stealth, scale=1.1]

  \begin{scope}[xshift=2.2cm]

    \draw[->] (0,0) -- (6.5,0) node[right] {\large QIV Levels};
    \draw[->] (0,0) -- (0,4.5) node[above] {\large $Y$};

    \node at (1.5,-0.3) {\normalsize $Z=0$};
    \node at (4.5,-0.3) {\normalsize $Z=1$};

    \draw[thick,red] (1.5,2.2) -- (5,3.2) node[anchor=west] {\footnotesize \textcolor{red}{$E(Y_1 \mid A=1,Z=1,X=x)$}};
    \draw[thick,blue] (1.5,1.2) -- (5,2.2) node[anchor=west] {\footnotesize \textcolor{blue}{$E(Y_0 \mid A=1,Z=1,X=x)$}};

    \filldraw[red] (1.5,2.2) circle (2pt);
    \filldraw[red] (5,3.2) circle (2pt);
    \filldraw[blue] (1.5,1.2) circle (2pt);
    \filldraw[blue] (5,2.2) circle (2pt);

    \node[anchor=east, text=red] at (-0.2,2.2) {\footnotesize $E(Y_1 \mid A=1,Z=0,X=x)$};
    \node[anchor=east, text=blue] at (-0.2,1.2) {\footnotesize $E(Y_0 \mid A=1,Z=0,X=x)$};

    \draw[dashed, red] (-0.2,2.2) -- (1.5,2.2);
    \draw[dashed, blue] (-0.2,1.2) -- (1.5,1.2);

    \draw [decorate,decoration={brace,amplitude=8pt}] (1.2,1.2) -- (1.2,2.2) node[midway,left=6pt] {\footnotesize $\gamma(x)$};

  \end{scope}
\end{tikzpicture}
\caption{
Illustration of no current treatment value interaction in a QIV framework with stratification by $Z$. The potential outcomes $E(Y_1 \mid A=1, Z, X=x)$ and $E(Y_0 \mid A=1, Z, X=x)$ for treated units ($A=1$) evolve in parallel across values of the QIV $Z$. The vertical gap $\gamma(x)$ represents the ATT at $Z=0$. The homogeneity assumption implies that $E(Y_0 \mid A=1, Z, X=x)$ follows the same trend as $E(Y_1 \mid A=1, Z, X=x)$ across $Z$ among treated subjects, justifying interpretation of $\gamma(x)$ as a conditional ATT that is constant across $Z$.}

\end{figure}

\begin{theorem}
If Assumptions \ref{assum3}-\ref{assum1} hold, then
the conditional confounding bias parameter  $\alpha(x)$ and the conditional ATT  $\gamma(x)$  are identified  for $z \in \{0,1\}$ as follows:
	\begin{eqnarray*}
	\alpha(x) &=&\frac{E(Y|A=1,Z=1, X=x)  -E(Y|A=1,Z=0,X=x)} {E(Y|A=0,Z=1,X=x)-E(Y|A=0,Z=0,X=x)},  \label{eqn:alpha_identi}\\\
	\gamma(x) &=& E(Y|A=1,Z=z,X=x) - \alpha(x) E(Y|A=0,Z=z,X=x). \label{eqn:gamma_identi}
	\end{eqnarray*}
\label{thm: nonpara-identi-multipli}
The marginal ATT $\gamma$ equals the average value of $\gamma(x)$ with respect to the distribution of $X$ among treated individuals: $\gamma = E_{X}\{\gamma(X)|A=1\}.$
\end{theorem}
The following remarks are warranted at this juncture. 
\begin{Remark}
    Revisiting the analogy with nonlinear difference-in-differences (DiD) with repeated cross-sectional sampling, the multiplicative parallel trends assumption is equivalently stated  $\frac{P \left\{ Y_{a=0}=1|A=1,Z=1\right\} }{P \left\{ Y_{a=0}=1|A=1,Z=0%
\right\} }=\frac{P \left\{ Y_{a=0}=1|A=0,Z=1\right\} }{P \left\{ Y_{a=0}=1|A=0,Z=0%
\right\} }$. 
From this perspective our approach may be viewed as a modified DiD approach to allow for violation of no anticipation, by replacing the latter condition with the no additive current treatment value interaction. Importantly, this now allows for $Z=T=0$ to occur post treatment assignment such that the treatment can impact the outcome causally at $T=0$, it also allows for $Z$ to potentially be any measured covariates, not necessarily encoding time, for which the conditions hold, arguably providing richer opportunities for identification than standard DiD. A variable designated to play this privileged role is what we have named a QIV.   
\end{Remark}
 
\begin{Remark}
The relationship between the additive confounding bias   $\delta(z,x)$ and the multiplicative confounding bias   $\alpha(x)$ is  $\delta(z,x)=\{\alpha(x)-1\}E(Y|A=0,Z=z,X=x)$ under Assumption \ref{assum2}. If $\alpha(x) =1$, then the confounding bias  vanishes; if $\alpha(x) > 1$, then the confounding bias is upward; and if $\alpha(x) <1$, then the confounding bias is downward.  It is important to note that the confounding bias  on the additive scale $\delta(z,x)$ varies with $z$, but the confounding bias  on the multiplicative scale $\alpha(x)$ remains constant with respect to $z$ when $\alpha(x) \neq 1$.
\end{Remark}

\begin{Remark}
 We have also obtained identification results for the causal risk ratio in the treated subpopulation, which are presented in Section S1 of the Supplementary Materials.
\end{Remark}

\subsection{Lack of Identification with Parallel Trends on the Odds Ratio Scale}

Interestingly, in the case of a continuous outcome, \citet{liu2022mendelian} considered the no current treatment value interaction Assumption \ref{assum1} in conjunction with an alternative parallel trends assumption on the odds ratio scale to establish identification of the additive ATT.  
In sharp contrast to the continuous outcome case, we show in Section S3 of the Supplementary Materials that, for a binary outcome, the additive ATT may not be uniquely identified if parallel trends is assumed to hold on the odds ratio scale rather than on the multiplicative scale as in Assumption~\ref{assum2}.

\section{ Bounded and Efficient Estimation  }
\label{sec:estimation}
\subsection{A Novel Nested Nuisance Model }
Our established identification results in Section \ref{sec:identify} allows us to estimate the marginal ATT $\gamma$ on the additive scale and the confounding bias $\alpha$ on the multiplicative scale. Let \( p_{aa^*}(z,x) = \Pr(Y_a = 1 \mid A = a^*, Z = z, X = x) \). These conditions naturally imply the following additive-multiplicative  model for the observed conditional  risk 
\begin{equation}
P \left( Y=1|A=a,Z=z,X=x\right) =\gamma(x) a+\alpha
^{a}(x)p_{00}\left( z,x\right),\label{eq:gamma_alpha}
\end{equation}
where $\left( \gamma(x),\alpha(x)\right) \in \left( -1,1\right)
	\times \mathbb{R}^{+}$, and the conditional observed baseline risk  is $p_{00}\left( z,x\right) = P \left( Y=1|A=0,Z=z,X=x\right)$.  Note that a similar additive--multiplicative model for the observed data has appeared previously in the survival analysis literature \citep{lin1995semiparametric}, albeit in a different non-causal context. 
    
We can observe \( p_{00}(z,x) \) and \( p_{11}(z,x) \), but not \( p_{01}(z,x) \).
Both $p_{01}(z,x)$ and $p_{00}(z,x)$ are probabilities and thus must lie in the unit interval $(0,1)$. As a result,  the baseline risk $p_{00}\left( z,x\right)$ is {\it variation dependent} on the parameters $\gamma(x)$ and $\alpha(x)$. For example, if $\gamma(x)=0.1$ and $\alpha(x)=1.2$, then $p_{00}(z,x)\leq 0.9/1.2 =0.75$ because  $ 0 \leq  p_{01}(z,x), p_{11}(z,x) \leq 1$.  Therefore, the joint ranges of $(\gamma(x),\alpha(x), p_{00}(z,x))$ are strictly smaller than the Cartesian product of the marginal ranges of $\gamma(x),\alpha(x)$ and $p_{00}(z,x)$. Consequently, a direct specification of  the baseline risk $p_{00}(z,x)$  will fail to work here because it cannot guarantee the probabilities of $p_{11}(z,x)$ and $p_{01}(z,x)$ to be within the unit interval.  To resolve this issue, we propose the  generalized odds product (GOP)  parameterization  by extending previous results 
\citep{Richardson2017,Yin2021} to specify the following nuisance model
\begin{equation}
	\mathrm{GOP}\left( z,x\right) =\frac{p_{11}\left( z,x\right) }{1-p_{11}\left(
	z,x\right) }\times \frac{p_{01}\left( z,x\right) }{1-p_{01}\left( z,x\right) }%
\times \frac{p_{00}\left( z,x\right) }{1-p_{00}\left( z,x\right) }\in \mathbb{R}^{+}.\label{eq:GOP_z}
\end{equation}
We obtain the following result showing that our proposed GOP nuisance model parameterization resolves the variation dependence issue. The proof of Theorem~\ref{thm:diffeo} is given in Section S4.2 of the Supplementary Materials.
\begin{theorem}
	\label{thm:diffeo}
	If $\gamma(x)$, $\alpha(x)$  and $\mathrm{GOP}\left( z,x\right)$ are defined in equations (\ref{eq:gamma_alpha}) and (\ref{eq:GOP_z}) respectively, then the following map 
	\[
	\left( p_{11}\left( z,x\right) ,p_{01}\left( z,x\right) ,p_{00}\left( z,x\right)
	\right) \rightarrow \left( \gamma(x),\alpha(x),
	\mathrm{GOP}\left( z,x\right) \right), 
	\]%
	is a diffeomorphism. That is, for any $\left( \gamma(x),\alpha(x),\mathrm{GOP}\left( z,x\right) \right) \in \left( -1,1\right)
	\times \mathbb{R}^{+}\times \mathbb{R}^{+},$ there is one and only one
	vector $\left( p_{11}\left( z,x\right), p_{01}\left( z,x\right) ,p_{00}\left(
	z,x\right) \right) \in \left( 0,1\right) ^{3}$. 
\end{theorem}
Theorem \ref{thm:diffeo} states that the domain of $\left( \gamma(x) ,\alpha(x) ,\mathrm{GOP}(z,x)\right) \in \left(
-1,1\right) \times \mathbb{R}^{+}\times \mathbb{R}^{+}$ is the Cartesian
product of the domains of $\gamma(x),\alpha(x)$ and $\mathrm{GOP}(z,x)$, and thus the
model parameters are {\it variation independent} \citep{tsiatis2007semiparametric}.  Let $b_1(z,x) = \{1 + \mathrm{GOP}(z,x)\}\alpha^2(x)$, $b_2(z,x)=\alpha(x)\gamma(x) - \mathrm{GOP}(z,x)\times \{\alpha^2(x) +2\alpha(x) -\alpha(x)\gamma(x)\}$, $b_3(z,x)=-\mathrm{GOP}(z,x)\times \{\alpha(x)\gamma(x) + \gamma(x) - 2\alpha(x) -1\}$, $b_4(z,x) = \mathrm{GOP}(z,x)\times \{\gamma(x)-1\}$. Given $\gamma(x),\alpha(x)$, and $\mathrm{GOP}(z,x)$, the baseline risk $p_{00}(z,x)$ satisfies the following cubic equation
\[
b_1(z,x)p^3_{00}(z,x) + b_2(z,x)p^2_{00}(z,x) + b_3(z,x)p_{00}(z,x) + b_4(z,x) =0.
\]

We show that this equation admits a unique root within the interval 
\[
\left(\max \left( 0,\frac{-\gamma(x)}{\alpha(x)} \right), 
      \min \left( \frac{1-\gamma(x)}{\alpha(x)}, \frac{1}{\alpha(x)}, 1 \right) \right),
\]
which is contained in the unit interval $(0,1)$. In practice, numerical methods can be employed to find the unique real root within the specified interval \citep{brent2013algorithms}. Alternatively, the baseline risk $p_{00}(z,x)$ has a closed-form solution derived from the theory of cubic equations \citep{abramowitz1965handbook}; the explicit derivation using Cardano's method is provided in Section~S4.3 of the Supplementary Materials.

\begin{figure}
    \includegraphics[width=1\linewidth]{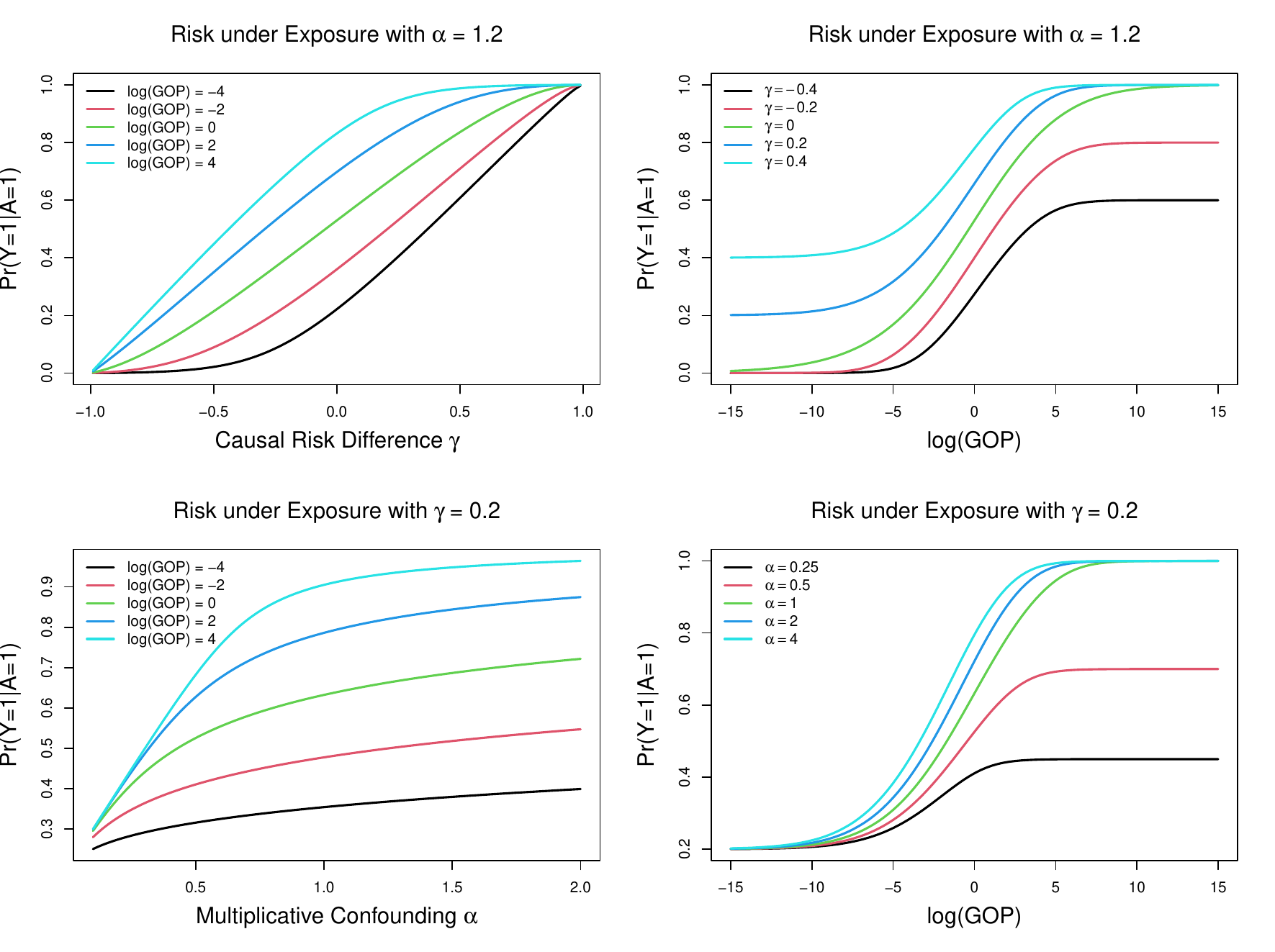}
    \caption{Implied probabilities from the proposed GOP parameterization.}
    \label{fig:GOP}
\end{figure}

This result implies that for any posited models for the conditional ATT $\gamma(x)$, the multiplicative confounding bias $\alpha(x)$ and $\mathrm{GOP}\left( z,x\right)$, one is guaranteed to uniquely specify $p_{00}(z,x)$,  $p_{01}(z,x) = \alpha(z,x)p_{00}(z,x)$ and $p_{11}(z,x) = \gamma(z,x)+\alpha(z,x)p_{00}(z,x)$ all which fall in the unit interval $(0,1)$.   To illustrate, Figure~\ref{fig:GOP} plots the disease risk under exposure $A = 1$ as a function of $\gamma$, $\alpha$, and $\log(\mathrm{GOP})$. We observe that the GOP parameterization respects the natural constraints on probabilities and ensures that the disease risk remains within the unit interval $(0,1)$. Notably, when the true causal risk difference $\gamma = 0.2$, our proposed model guarantees that the disease risk among the exposed lies within $(0.2, 1)$, since the counterfactual risk (i.e., the risk if everyone were unexposed) among the exposed cannot be negative. Therefore, the proposed GOP nuisance model provides a flexible and principled framework for modeling risk functions, ensuring both mathematical coherence and compatibility with causal interpretations.

\subsection{Maximum Likelihood Estimation}
\label{sec:MLE}
Suppose that $({O}_1,...,{O}_n)$ are independent and identically distributed (i.i.d.) observations of $O=(Y,A,Z,X)$, and let  $P_n$ denote the empirical measure, $P_n\{h(O)\} = n^{-1}\sum_{i=1}^n h(O_i)$ for some square-integrable and measurable function $h(\cdot)$. 
We first describe a principled maximum likelihood estimation (MLE) approach under parametric models for the conditional ATT $\gamma({x};{\beta})$, the conditional confounding bias  $\alpha({x};\theta{})$, and the GOP model conditional on  covariates $X=x$:
\begin{eqnarray}
	\gamma({x};{\beta}) &=&\tanh \left( \beta^Tx \right) =\frac{\exp \left(
	2\beta^Tx \right) -1}{\exp \left( 2\beta^Tx  \right) +1} \in (-1,1), \label{eqn:gamma-model}  \nonumber\\
		 \alpha({x};\theta{}) &=&\exp \left( \theta^T x\right) >0, \label{eqn:alpha-model}  \nonumber \\
\mathrm{GOP}(z,{x};\omega_0,\omega_1,\eta) & =& \exp(\omega_0 + \omega_1z + {\eta}^T x) >0.
\nonumber
\end{eqnarray}

Under our proposed parametrization,  the conditional distribution of the outcome $Y$ is given by
 \[
 Y|A,Z,{X}\sim \mathrm{%
 	Bernoulli}\left( P(Y=1\mid A,Z,X;\varphi)\right),\quad \varphi=(\beta^T ,\theta^T ,\omega_0,\omega_1, {\eta}^T),
 \]
 where $P(Y=1\mid A,Z,X;\varphi)=\tanh \left( \beta \right) A$+$\exp \left( \theta A\right) p_{00}\left(
 Z,{X};\beta ,\theta ,\omega_0,\omega_1, {\boldeta} \right)$,
 and the baseline risk $p_{00}\left(
 Z,{X};\beta ,\theta ,\omega_0,\omega_1, {\eta} \right) $ can be obtained numerically using a univariate root-finding algorithm, such as the bisection method \citep{brent2013algorithms}. 
The log-likelihood is thus given by 
\begin{equation}
\ell_n(\varphi) = nP_n\{\ell(O;\varphi)\},
\label{eqn:loglik}
\end{equation}
where $\ell(O;\varphi)=Y\log(P(Y=1\mid A,Z,X;\varphi))$+$(1-Y)\log\{1-P(Y=1\mid A,Z,X;\varphi)\}$, which can be maximized  directly using a standard unconstrained optimization method to obtain parameter estimates. Standard errors and confidence intervals for the likelihood parameters can also be obtained in standard fashion, and inference about the marginal ATT follows directly from the maximum likelihood plug-in principle.  

The proposed maximum likelihood estimator has several notable advantages. First, the parametrization is perfectly coherent with model restrictions, so that the model is guaranteed to produce predictions that are within their natural range. Second, the estimator attains the efficiency bound for the posited model in large samples, which is an important benefit, particularly given well known efficiency challenges of IV analyses. Also, because the joint distribution for the treatment and candidate instrument given covariates factorizes, the estimator does not require modeling the latter and is therefore robust to any bias due to misspecification of such a law. Finally, a straightforward score test, or a likelihood ratio test can readily be implemented to test the null hypothesis $H_0$ of no conditional ATT. As noted earlier, the approach is guaranteed to have correct type 1 error provided Assumptions \ref{assum3} and \ref{assum2} hold, as Assumption \ref{assum1} holds automatically under $H_0$, as long as the likelihood is correctly specified. Although, we expect such a test will be guaranteed to be consistent only against alternatives in a model where Assumption \ref{assum1} holds, and might have trivial power outside the model.       
Despite several appealing properties, the MLE does require correct specification of the outcome law, and therefore may be susceptible to bias in case the model is wrong. In the next section we propose a semiparametric estimator with appealing robustness property. 

\subsection{Semiparametric Locally Efficient Triply Robust Estimation}
\label{sec:DR.est}
Often, one may not be certain that the posited baseline outcome model is correctly specified.  Therefore, we consider an alternative potentially more robust estimator to partially guard against model misspecification. 
In this vein, we now propose an estimator that is consistent for the causal effect even if the outcome regression model $p_{00}(x,z)$ is not correctly specified, provided that working models for $\alpha({x})$ and
\begin{equation*}
P(A,Z|{X}) = P(A|Z,{X})P(Z|{X}) ,    
\end{equation*}
are correctly specified. Note that  modeling  $P(A,Z|{X})$ does not involve the outcome $Y$, and thus can be carried out separately from the MLE using only data on $(A,Z,X)$. In order to obtain the desired triply robust estimator, we derive the efficient influence function (EIF) of  $\gamma$  in the nonparametric model in which no prior restriction is placed on the observed data law \citep{newey1990semiparametric, bickel1993efficient}. 
\begin{theorem}
The efficient influence function  $\psi(O;\gamma)$ for estimating $\gamma$ in the nonparametric model for the distribution of $O=(Y,A,Z,{X})$ is given by
\begin{eqnarray*}
&& \frac{1}{P(A=1)}\Biggr(P\left( A=1|X,Z\right) \frac{\left\{ YA-Y(1-A)\alpha \left( X\right)
\right\} }{P\left( A|Z,X\right) } +\gamma \left( X\right) \left( A-P\left( A=1|X,Z\right) \right)  \\
&&-\left[ \frac{P\left( A=1|X,Z\right) \left\{ E\left( Y|A,X,Z\right)
A-E\left( Y|A,X,Z\right) (1-A)\alpha \left( X\right) \right\} }{P\left(
A|Z,X\right) }-\gamma \left( X\right) P\left( A=1|X,Z\right) \right]  \\
&&-E\left\{ P\left( A=1|X,Z\right) \left[ E\left( Y|A=0,X,Z\right) \right]
|X\right\}  \\
&&\times \left\{ \frac{\left\{ Y-E\left( Y|A,Z,X\right) \right\} }{\left[
E\left( Y|A=0,Z=1,X\right) -E\left( Y|A=0,Z=0,X\right) \right] }\frac{\left(
-1\right) ^{A+Z}}{P\left( A,Z|X\right) }\alpha \left( X\right)
^{1-A}\right\} \Biggr)-\gamma
\end{eqnarray*}
Accordingly, the semiparametric efficiency bound for estimating $\gamma$ is $E\{\psi^2(O;\gamma)\}$.
\label{thm:efficientIF}
\end{theorem}

The proof of Theorem~\ref{thm:efficientIF} is provided in Section S4.3 of the Supplementary Materials. To establish triply robustness of the efficient  influence function, consider the following three semiparametric models $\mathcal{M}_1$, $\mathcal{M}_2$, and $\mathcal{M}_3$:
\begin{description}
\item[$\mathcal{M}_1$]:  $P \left( Y=1|A=0,X,Z\right) ,\gamma \left( X\right) ,\alpha
\left( X\right) $  are correctly specified parametric models while the rest of the model is unrestricted;
\item[$\mathcal{M}_2$]:   $\alpha \left( X\right) $ and $P\left( A,Z|X\right) $ are correctly specified while the rest of the model is unrestricted;
\item[$\mathcal{M}_3$]: $\gamma \left( X\right) ,P \left( Y=1|A=0,X,Z\right) $ and $P\left(
A,Z|X\right) $ are correctly specified while the rest of the model is unrestricted.
\end{description}

In the proof of Theorem~\ref{thm:tr.ATT} (see Section S4.4 of the Supplementary Materials), 
we establish that the moment condition $E\{\psi(O;\gamma)\}=0$ holds in the union model 
$\mathcal{M}_1 \cup \mathcal{M}_2 \cup \mathcal{M}_3$. Under standard regularity conditions, 
this implies that the corresponding moment-based estimator for the marginal ATT $\gamma$ 
is consistent when solving the empirical version of the moment condition,

\begin{equation}
\label{eqn:IF.ATT}
\begin{aligned}
&\widehat{\gamma}_{tr}=\frac{1}{\widehat{P}(A=1)} \Bigg\{ \widehat{P}\left( A=1|X,Z\right) \frac{\left\{ YA - Y(1-A) \widehat{\alpha} \left( X\right) \right\} }{\widehat{P}\left( A|Z,X\right) } + \widehat{\gamma} \left( X\right) \left( A- \widehat{P} \left( A=1|X,Z\right) \right) \\
&\quad -\left[\frac{\widehat{P}\left( A=1|X,Z\right) \left\{ \widehat{E}\left( Y|A,X,Z\right) A - \widehat{E}\left( Y|A,X,Z\right) (1-A) \widehat{\alpha} \left( X\right) \right\} }{\widehat{P}\left( A|Z,X\right) } - \widehat{\gamma} \left( X\right) \widehat{P}\left( A=1|X,Z\right) \right]  \\   
&\quad -\widehat{E}\left\{ \widehat{P}\left( A=1|X,Z\right) \left[ \widehat{E}\left( Y|A=0,X,Z\right) \right]
|X\right\}  \\
&\quad \times  \frac{\left\{ Y-\widehat{E}\left( Y|A,Z,X\right) \right\} }{\left[
\widehat{E}\left( Y|A=0,Z=1,X\right) -\widehat{E}\left( Y|A=0,Z=0,X\right) \right] } \frac{\left(
-1\right) ^{A+Z}}{\widehat{P}\left( A,Z|X\right) } \widehat{\alpha} \left( X\right)^{1-A}  \Bigg\},
\end{aligned}
\end{equation}
can be shown to be consistent and asymptotically normal (CAN) if either $\mathcal{M}_1$ or $\mathcal{M}_2$ or $\mathcal{M}_3$ holds. Such an estimator is therefore {\it triply robust}  as was sought out. 

Note that the triply robust estimator $\widehat{\gamma}_{tr}$ defined in equation~\eqref{eqn:IF.ATT} requires estimation of parameters from the following models: $\gamma(X)$, $\alpha(X)$, $P(Y = 1 \mid A = 0, X, Z)$, and $P(A, Z \mid X)$. The covariate vector $X$ is typically multivariate and may include continuous variables such as age. To reduce dimensionality and ensure stable estimation, we follow standard practice and specify parametric models for these nuisance components \citep{Robins1997}.

To estimate the weight function $P(A, Z \mid X)$, we factorize the joint distribution as $P(A \mid Z, X) P(Z \mid X)$, and fit logistic regression models for both components. That is, we fit two binary generalized linear models (GLMs) with logit link: one for $P(Z \mid X)$ and another for $P(A \mid Z, X)$.

For the conditional mean function $E(Y \mid A, Z, X)$ and its submodel $E(Y \mid A = 0, Z, X)$, we adopt maximum likelihood estimation under the proposed generalized odds product (GOP) parameterization. This formulation guarantees variation independence among nuisance parameters and enforces compatibility with the natural support of the binary outcome. Specifically, the baseline risk function $p_{00}(Z, X) = P(Y = 1 \mid A = 0, Z, X)$ is estimated by solving a cubic equation derived from the GOP parameterization, as described in Section~3.1. This equation admits closed-form roots and can be solved either analytically using Cardano’s formula or numerically.

We also estimate $\alpha(X)$ and $\gamma(X)$ using doubly robust estimating equations. Suppose that $\alpha(X)$ is correctly specified and either $\mathcal{M}_1$ or $\mathcal{M}_2$ holds. Then the following moment condition provides a consistent and doubly robust estimator for $\alpha(X)$:
\[
E\left[ h(X) \left\{ Z - P(Z = 1 \mid X) \right\}
\left\{ 
\frac{Y(1 - A)\alpha(X) - Y A}{P(A \mid Z, X)} - \gamma(X)
\right\} \right] = 0.
\]
This moment equation can also be used to construct a doubly robust score-like test of the causal null hypothesis $H_0$ of no conditional ATT, which will have nominal type 1 error under Assumptions \ref{assum3} and \ref{assum2}, and the union of models $\mathcal{M}_1$ and $\mathcal{M}_2$. Similarly, under the assumption that $\gamma(X)$ is correctly specified and either $\mathcal{M}_1$ or $\mathcal{M}_3$ holds, the following moment equation provides a doubly robust estimator for $\gamma(X)$:
\[
E\left[ h(X) \left\{ Z - P(Z = 1 \mid X) \right\}
\left\{
\frac{Y A - \gamma(X) A}{P(A \mid Z, X)P(Y = 1 \mid A = 0, X, Z)} - \alpha(X)
\right\} \right] = 0.
\]
Here, $h(X)$ is a user-specified (possibly vector-valued) function of the covariates $X$ that serves to index the estimating equations. All nuisance functions are estimated using parametric models by default, although the proposed framework can accommodate more flexible, nonparametric or machine learning-based approaches. The final estimator $\widehat{\gamma}_{tr}$ is obtained by plugging in the estimated nuisance quantities into the efficient influence function and computing the empirical average across the observed data.

Theorem~\ref{thm:tr.ATT} summarizes the properties of our proposed triply robust estimator $\widehat{\gamma}_{tr}$. The proof is given in Section S4.4 of the Supplementary Materials. 
\begin{theorem}
Under standard regularity conditions and the positivity assumption that $P(A,Z|{X})$ is bounded away from zero almost surely, $\widehat{\gamma}_{tr}$ is a CAN estimator in the union model $\mathcal{M}_1\cup \mathcal{M}_2\cup \mathcal{M}_3$.  Furthermore, $\widehat{\gamma}_{tr}$ attains the semiparametric efficiency bound in the nonparametric observed data model at the intersection sub-model $\mathcal{M}_1\cap \mathcal{M}_2 \cap \mathcal{M}_3$ where all the working models are correctly specified.
\label{thm:tr.ATT}
\end{theorem}

\subsection{Many Weak Quasi-Instruments}

\label{sec:weakIV}

Multiple QIVs can be incorporated under the proposed framework to improve identification and efficiency. For example, in an epidemiological analysis using genetic variants as QIV in the maximum likelihood estimation approach in Section \ref{sec:MLE} where $m$ genetic variants are available, one can replace the scalar $Z$ in the log-likelihood \eqref{eqn:loglik} by a $m$-dimensional vector $\boldsymbol{Z}$. Let $ \hat{\varphi} $ be the solution to the corresponding score function, i.e., $S_n( {\varphi})=\partial \ell_n(\varphi)/\partial \varphi=0$; and let   $$ 
I_1(\varphi)=  -E\left\{ \frac{\partial^2}{\partial \varphi\partial \varphi^T} \ell(O_1;\varphi)\right\}
$$ 
be the corresponding Fisher information matrix  for one observation. Let $k$ be the total number of parameters in $\varphi$.  When $\lambda_{\min}  \{n I_1(\varphi)\}/k \rightarrow\infty$ with  $ \lambda_{\min} \{n I_1(\varphi)\} $ being the minimum eigenvalue of $ n I_1(\varphi) $, according to the results in \citet{liu2022mendelian}, $\hat\varphi$ is consistent and asymptotically normal as follows
\begin{eqnarray}
\sqrt{n} (\hat{\varphi}- \varphi) \xrightarrow{d} N\big(0, \{ I_1 (\varphi)\}^{-1} \big), \label{eq: mle normality}
\nonumber 
\end{eqnarray}
as the sample size $n \rightarrow +\infty$.
In other words, the MLE is consistent and asymptotically normal if the total identification signal is strong enough. In practice, we can empirically evaluate this condition using the ratio between the minimum eigenvalue of the negative Hessian matrix and the number of parameters: 
$\hat\kappa= -\lambda_{\min} \{ I_n (\hat \varphi)\}/k$. In practice, we recommend checking that the $\hat{\kappa}$ is at least greater than 10  \citep{liu2022mendelian}. For the semiparametric estimation in Section \ref{sec:DR.est}, a fixed number of semiparametric moments can be incorporated
by adopting a standard generalized method of moments (GMM) approach \citep{hansen1982large}. We leave it as future work to consider the setting where the number of moments is allowed to diverge. Further technical details on the extension to settings with many (possibly weak) QIVs are provided in Section~S4.5 of the Supplementary Materials.

\section{Simulation Studies}
\label{sec:simulation}

In this section, we conduct simulation studies to evaluate the finite-sample performance of the proposed estimators for the marginal average treatment effect on the treated, denoted \( \gamma \), using both maximum likelihood estimation (denoted as MLE) and semiparametric triply robust (denoted as TR) estimation methods.

\subsection*{Data Generating Process}

We simulated data for \( n = 50{,}000 \) individuals. Each subject had two baseline covariates: a binary variable \( X_1 \sim \text{Bernoulli}(0.5) \) and a continuous variable \( X_2 \sim \mathcal{N}(0,1) \). These covariates jointly determined a binary genetic variant \( Z \sim \text{Bernoulli}(\text{logit}^{-1}(-0.5 + 0.2 X_1 - 0.1 X_2)) \). The treatment indicator \( A \sim \text{Bernoulli}(\text{logit}^{-1}(-0.2 + 0.1 Z - 0.1 X_1 + 0.05 X_2)) \) was generated based on both \( Z \) and \( X \), reflecting potential confounding.

The generalized odds product (GOP) term, capturing joint dependence between \( A \) and \( Y \), was generated via the log-linear form:
\[
\log(\text{GOP}) = \omega_0 + \omega_1 Z + \eta_1 X_1 + \eta_2 X_2,
\]
with \( \omega_0 = -5 \), \( \omega_1 = 3.5 \), \( \eta_1 = 1.5 \), and \( \eta_2 = 0.5 \). This specification explicitly models how \( X_1 \) and \( X_2 \) jointly influence the GOP along with the QIV \( Z \). We generated individual-level functions \( \gamma(X) \) and \( \alpha(X) \) as:
\begin{align*}
\gamma(X) &= \tanh(\beta_0 + \beta_1 X_1 + \beta_2 X_2), \\
\alpha(X) &= \exp(\theta_0 + \theta_1 X_1 + \theta_2 X_2),
\end{align*}
with fixed coefficients \( \beta = (0.3, 0.1, 0.1)^\top \) and \( \theta = (0.4, 0.2, 0.1)^\top \).

The observed outcome \( Y \) followed a Bernoulli distribution with event probability:
\[
P(Y = 1 \mid A, X, Z) = \gamma(X) \cdot A + \alpha(X)^A \cdot p_{00}(X, Z),
\]
where \( p_{00}(X, Z) \) was numerically solved to ensure that the outcome probabilities remained in the unit interval \( (0,1) \). The true ATT was defined as the average of \( \gamma(X) \) over the distribution of \( X \) among treated individuals, whose true value is 0.334. 

\subsection*{Model Misspecification Scenarios}

To assess the triply robustness of our proposed estimator as in (\ref{eqn:IF.ATT}), we consider scenarios in which the models in \( \mathcal{M}_1 \), \( \mathcal{M}_2 \), or \( \mathcal{M}_3 \) are misspecified. In these cases, instead of using \( X_2 \sim \mathcal{N}(0,1) \), the data analyst uses covariates \( X_2^* \) that are generated from a uniform distribution on the interval \( (-1,1) \). Specifically, we consider the following four settings:

\begin{description}
\item[All correct:] \( X_1, X_2 \) are used in models \( \mathcal{M}_1 \), \( \mathcal{M}_2 \), and \( \mathcal{M}_3 \);
\item[\( \mathcal{M}_1 \) correct:] \( X_1, X_2 \) are used in \( \mathcal{M}_1 \), but \( X_1, X_2^* \) are used in \( P(A, Z \mid X_1, X_2^*) \);
\item[\( \mathcal{M}_2 \) correct:] \( X_1, X_2 \) are used in \( \mathcal{M}_2 \), but \( X_1, X_2^* \) are used in \( \gamma(X_1, X_2^*) \);
\item[\( \mathcal{M}_3 \) correct:] \( X_1, X_2 \) are used in \( \mathcal{M}_3 \), but \( X_1, X_2^* \) are used in the model for \( \alpha(X_1, X_2^*) \).
\end{description}

\subsection*{Simulation Results}

\begin{figure}[H]
	\centering
	\includegraphics[scale=0.65]{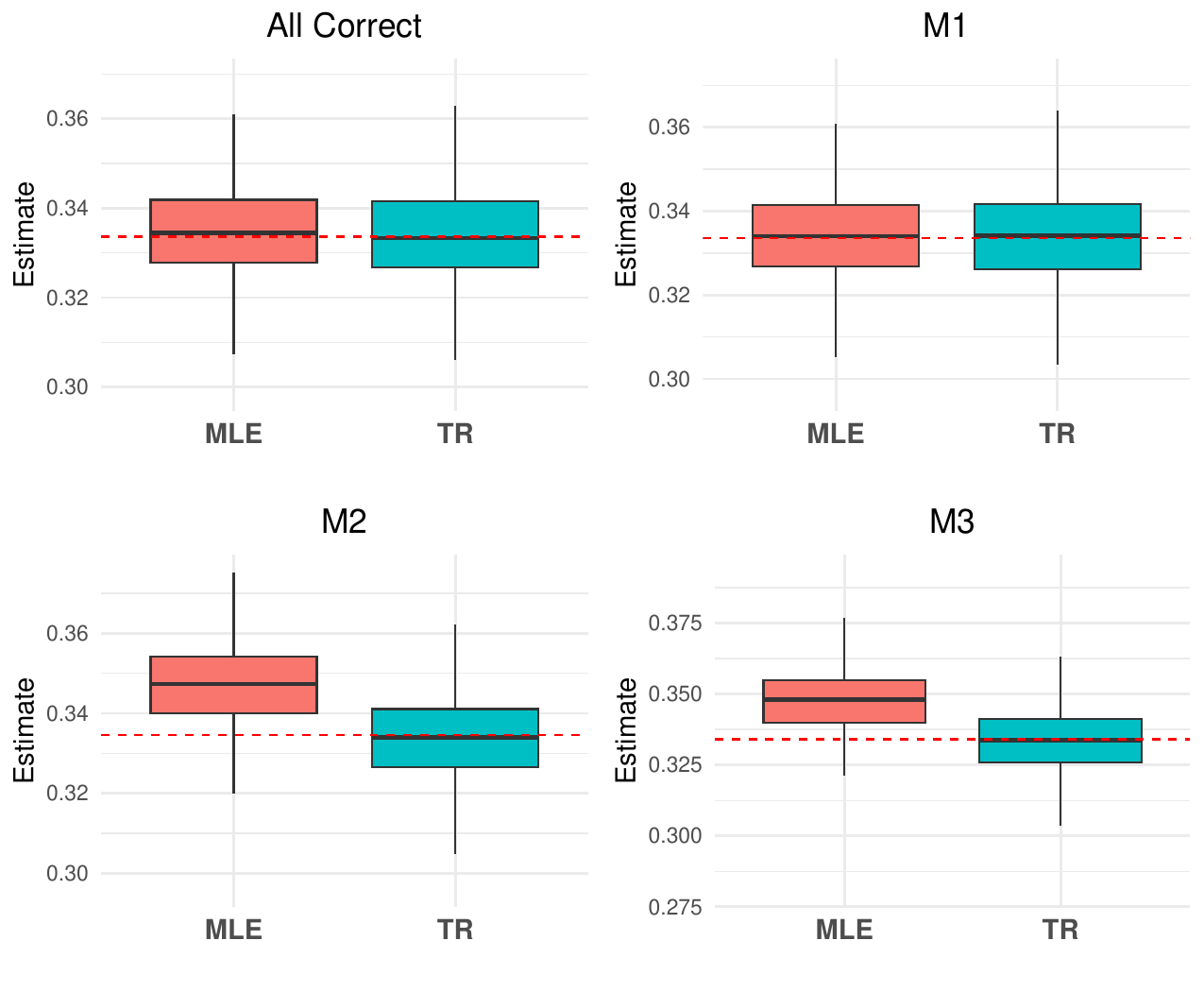}
	\caption{Boxplots comparing two estimators of the average treatment effect on the treated (ATT) — the maximum likelihood estimator (MLE) and the triply robust estimator (TR) — across four simulation settings: All Correct, $\mathcal{M}_1$, $\mathcal{M}_2$, and $\mathcal{M}_3$. In each panel, the red dashed line indicates the  true ATT 0.334.    }
	\label{fig:sim_boxplots}
\end{figure}

We conducted 1{,}000 simulation replicates under the four scenarios described above. In each replicate, we estimated: (1) the marginal ATT using a maximum likelihood estimator (MLE), and (2) the marginal ATT using our proposed {triply robust estimator} as defined in equation~\eqref{eqn:IF.ATT}. The performance of the proposed estimator across the different settings is summarized in Figure~\ref{fig:sim_boxplots}.

The results demonstrate that the proposed triply robust estimator is consistent across all four scenarios, confirming its triple robustness property: it yields consistent estimation as long as at least one of the three working models, \( \mathcal{M}_1 \), \( \mathcal{M}_2 \), or \( \mathcal{M}_3 \), is correctly specified. In contrast, the MLE is biased in either \( \mathcal{M}_2 \) or \( \mathcal{M}_3 \), as expected, since MLE relies on correct specification of the outcome model. However, when all models are correctly specified---or under scenario \( \mathcal{M}_1 \), where the outcome model is correctly specified---the MLE also provides consistent estimates, owing to correct specification of the likelihood for the outcome.

\section{Real Data Analysis: Adiposity and Hypertension in the UK Biobank}
\label{sec:ukb}

We applied our proposed QIV framework to evaluate the causal effect of being overweight on the risk of hypertension in the UK Biobank (UKB). The UKB is a large-scale prospective cohort study of over 500{,}000 individuals aged 40--69 years, recruited between 2006 and 2010 across 22 assessment centers in the United Kingdom. Participants contributed detailed phenotypic and genotypic information, including health questionnaires, physical measurements (e.g., height, weight, blood pressure), biospecimens (e.g., blood, urine, saliva), and linked electronic health records \citep{sudlow2015uk, bycroft2018uk}. For this analysis, we restricted the sample to 309{,}528 unrelated individuals of white British ancestry with complete measurements on body mass index (BMI), systolic blood pressure (SBP), and diastolic blood pressure (DBP).

The exposure of interest was a binary indicator of intermediate adiposity (overweight and obese), defined as a BMI greater than 25\,kg/m\textsuperscript{2} \citep{clarke2015estimating, timpson2009does}. The outcome was hypertension, defined as SBP $\geq$ 140\,mmHg or DBP $\geq$ 90\,mmHg. Among eligible individuals, the mean age was 57 years (SD = 7.98), 46.3\% were male, and the average BMI was 27.37\,kg/m\textsuperscript{2}. Approximately 66.6\% were classified as overweight, and 51.2\% met criteria for hypertension.

We first estimated the marginal average treatment effect on the treated (ATT) using the parametric g-formula. Specifically, we fit a series of logistic regression models for the binary outcome of hypertension, regressing the outcome on treatment and selected baseline covariates, and computed counterfactual risks by averaging predicted probabilities under observed and counterfactual treatment assignments. The unadjusted ATT was 0.1807 (95\% CI: 0.1771, 0.1844). Adjustment for sex reduced the ATT to 0.1643 (95\% CI: 0.1607, 0.1686), and further adjustment for age, smoking history, and socioeconomic status (Townsend deprivation index) attenuated the estimate to 0.1478 (95\% CI: 0.1442, 0.1519). These results highlight the role of measured confounding in inflating the overweight--hypertension association.

\begin{figure}[ht]
    \centering
    \includegraphics[width=1.0\textwidth]{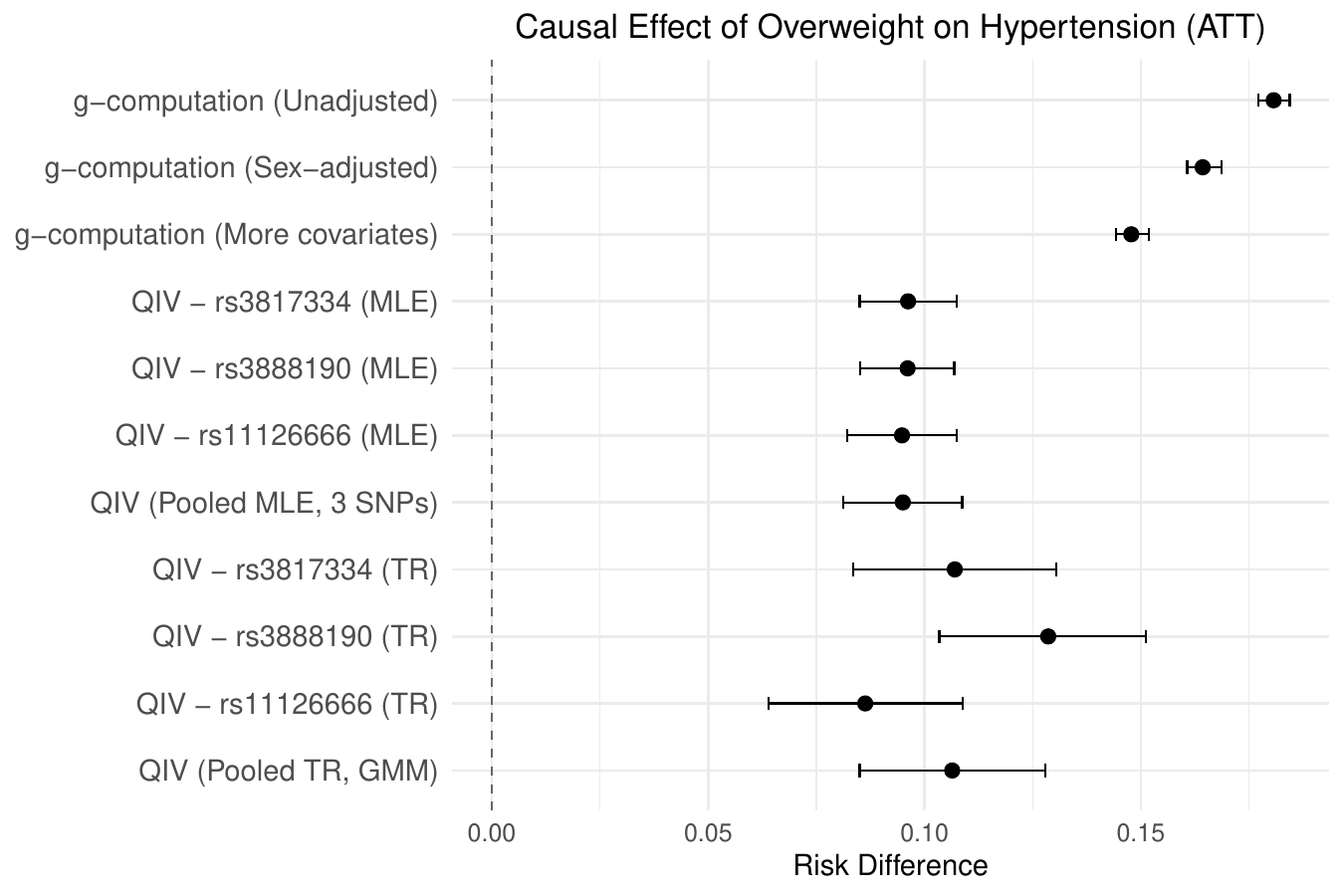}
    \caption{
        Forest plot of ATT estimates for the effect of overweight on hypertension risk using the g-computation and QIV methods. Error bars represent 95\% confidence intervals. The g-computation results (unadjusted, sex-adjusted, and more covariates) are presented first, followed by SNP-specific ATT estimates using the QIV maximum likelihood estimator (MLE) and triply robust (TR) estimator. All QIV analyses are adjusted for age and sex.
    }
    \label{fig:forest}
\end{figure}

The three SNPs---\texttt{rs3817334}, \texttt{rs3888190}, and \texttt{rs11126666}---were selected from a previously published Mendelian randomization study of adiposity and cardiometabolic risk \citep{sun2019type}, based on their documented relevance to adiposity-related pathways. 
To investigate the potential impact of residual unmeasured confounding, we use these SNPs as QIVs. For each SNP, we estimated the average treatment effect on the treated (ATT) using both the maximum likelihood estimator (MLE) and the proposed triply robust (TR) estimator, adjusting for sex and age. The MLE estimates were tightly clustered across SNPs, ranging from 0.0948 to 0.0962, reflecting stability under the parametric modeling assumptions. In contrast, the TR estimates ranged from 0.0863 to 0.1286, exhibiting moderate variability across SNPs. This variation is expected given that the TR estimator integrates information from multiple nuisance models and remains consistent under broader conditions, offering protection against model misspecification. Notably, all TR estimates were lower than that of the fully adjusted g-computation, suggesting that traditional outcome regression methods may still be subject to unmeasured confounding. Across all three SNPs, the marginal $\alpha$ estimates were consistently close to 1.14, indicating modest upward confounding---i.e., a tendency for conventional estimates to overstate the causal effect of overweight on hypertension risk. Figure~\ref{fig:forest} summarizes the point estimates and confidence intervals for the ATT estimates obtained from the g-computation and QIV methods.

The differences between the MLE and TR estimators reflect their distinct strengths and underlying assumptions. The MLE is fully efficient when the parametric model for the outcome is correctly specified, but it is sensitive to model misspecification. In contrast, the TR estimator is \emph{triply robust}: it remains consistent and asymptotically normal if any one of three key semiparametric models is correctly specified, as detailed in Section~\ref{sec:DR.est}. This robustness makes the TR estimator particularly well-suited to complex observational settings where model assumptions may be partially violated. The somewhat wider variation in TR estimates across SNPs likely reflects its reliance on multiple nuisance function estimates and its responsiveness to local differences in instrument strength---an expected trade-off for enhanced robustness. Consequently, moderate differences between TR and MLE estimates in finite samples are natural and do not indicate contradiction. Rather, they highlight the balance between efficiency (MLE) and robustness (TR). The general agreement between the two estimators in direction and scale reinforces the validity of our findings, while any divergence offers a useful diagnostic for evaluating potential model misspecification and residual confounding.

Importantly, the TR estimates were \emph{consistently lower than that of the adjusted g-computation with more covariates}, providing evidence that residual confounding, uncorrected even after covariate adjustment, may still inflate standard regression-based estimates. As shown in Figure~7, both the pooled TR (GMM) and pooled MLE estimates 
provide consistent evidence of a positive causal effect. For comparison, we also examined the classical two-stage least squares (2SLS) estimator using all three SNPs jointly. 
The 2SLS point estimate (0.173, 95\% CI: $[-0.014, 0.360]$) was close to the unadjusted g-computation estimate, 
suggesting that it likely did not remove unmeasured confounding. 
Moreover, the wide interval covering the null reflects the weakness of these instruments under the conventional IV framework. 
It is important to note that our three selected SNPs were chosen under the QIV framework, 
not the classical IV assumptions, and therefore may fail to meet IV relevance or other core IV conditions. 
Our proposed QIV framework, by leveraging genetic variants and semiparametric estimation, provides a credible complementary approach to causal inference in addition to standard adjustment methods or 2SLS. 
Additional empirical results, including two-stage least squares (2SLS) estimates, are provided in Section~S5 of the Supplementary Materials.

\section{Discussion}
\label{sec:discussion}

This paper presents a general and robust framework for causal inference that leverages QIV  to identify and estimate the average treatment effect on the treated  in the presence of unmeasured confounding. Classical IV approaches rely on strong assumptions such as independence from unmeasured confounders and the exclusion restriction, both of which may be routinely violated in observational studies. In contrast, our QIV approach enables point identification under a new set of empirically grounded stability conditions.

A fundamental contribution of this work is the development of a general identification strategy justified under a structural equilibrium generative model with hidden confounding. Such a model implies stability of the additive treatment effect across levels of a variable predictive of the outcome among the untreated (i.e., a 'quasi-instrument') and stability of confounding bias on the multiplicative scale.

Although our methodological development and empirical illustration focus on binary outcomes, the proposed identification strategy is not limited to this setting. It generalizes to continuous and time-to-event outcomes, and lends itself to applications in complex longitudinal studies, where exposures and confounders evolve over time.
Looking ahead, the QIV framework opens the door to numerous methodological extensions. These include adaptation to survival and longitudinal outcomes, incorporation of many or high-dimensional genetic instruments, and the integration of machine learning techniques for flexible nuisance function estimation. More broadly, the identification strategy proposed here provides a foundational contribution to causal inference---extending the reach of quasi-instrumental approaches to settings where traditional IV assumptions are untenable and enabling more robust and generalizable conclusions in complex observational studies.

\section*{Acknowledgment}
This research has been conducted using the UK Biobank Resource under application number 52008.

\clearpage
\renewcommand{\thesection}{S\arabic{section}}
\renewcommand{\thesubsection}{S\arabic{section}.\arabic{subsection}}
\setcounter{section}{0}

\section*{Supplementary Materials}
\addcontentsline{toc}{section}{Supplementary Materials}

\setcounter{theorem}{0}
\renewcommand{\thetheorem}{S\arabic{theorem}}

\setcounter{lemma}{0}
\renewcommand{\thelemma}{S\arabic{lemma}}

\setcounter{corollary}{0}
\renewcommand{\thecorollary}{S\arabic{corollary}}

\setcounter{assumption}{0}
\renewcommand{\theassumption}{S\arabic{assumption}}

\setcounter{example}{0}
\renewcommand{\theexample}{S\arabic{example}}

\setcounter{definition}{0}
\renewcommand{\thedefinition}{S\arabic{definition}}

\setcounter{figure}{0}
\renewcommand{\thefigure}{S\arabic{figure}}

\setcounter{table}{0}
\renewcommand{\thetable}{S\arabic{table}}

\setcounter{equation}{0}
\renewcommand{\theequation}{S\arabic{equation}}

\section{More Identification Results}
In this section, we present two novel identification results when the causal effect is defined on the multiplicative scale (causal risk ratio), and the confounding bias is defined on the additive scale or odds ratio scale. 

\subsection{Multiplicative Causal Effect and Additive Confounding Bias}
If the causal effect is defined  on the multiplicative scale and the confounding bias is defined on the additive scale as follows
\begin{eqnarray*}
\tilde{\gamma} &=& \frac{E(Y\mid A=1)}{E(Y_0\mid A=1)} \in \mathbb{R}^+,\\
\tilde{\alpha} &=& E(Y_0\mid A=1) - E(Y\mid A=0) \in (-1,1).
\end{eqnarray*}
Then, with a possibly invalid  instrument $Z$, we stratify the population by the value of $Z$ to obtain conditional causal effect $\tilde{\gamma}(z)$ and conditional confounding bias $\tilde{\alpha}(z)$ as follows 
\begin{eqnarray*}
\tilde{\gamma}(z) &=& \frac{E(Y\mid A=1,z)}{E(Y_0\mid A=1,z)} \in \mathbb{R}^+,\\
\tilde{\alpha}(z) &=& E(Y_0\mid A=1,z) - E(Y\mid A=0,z) \in (-1,1).
\end{eqnarray*}
 We obtain the following identification result. 
 
\begin{theorem}
Under the following three assumptions: (a) the causal effect $\tilde{\gamma}(z)$ does not vary with $Z$; (b) the confounding bias $\tilde{\alpha}(z)$ does not vary with $Z$; and (c) the $E[Y|A=1,Z=z]$ is not a constant function in $z$. Then, the causal effect $\tilde{\gamma}$ and confounding bias $\tilde{\alpha}$ are identified as follows:
\begin{eqnarray*}
\tilde{\gamma} &=& \frac{E(Y|A=1,Z=z)}{\tilde{\alpha}+E(Y|A=0,Z=z)}, \\
\tilde{\alpha} &=& \frac{E(Y|A=1,Z=1)E(Y|A=0,Z=0)-E(Y|A=1,Z=0)E(Y|A=0,Z=1)}{E(Y|A=1,Z=0)-E(Y|A=1,Z=1)}.
\end{eqnarray*}
\end{theorem}

 Although this multiplicative-additive model can be nonparametrically identified, however, this model has intrinsic drawbacks such that model parameter estimation is challenging.   To ensure the probability $E(Y|A=1,z)$ to be within the unit interval, that is  $0\leq \tilde{\gamma} \tilde{\alpha} + \tilde{\gamma}E(Y|A=0,z) \leq 1$, then $\tilde{\gamma} \tilde{\alpha} \leq 1$. This implies that the causal effect $\tilde{\gamma}$ and the confounding bias $\tilde{\alpha}$ cannot be variation independent \citep{tsiatis2007semiparametric}. For this reason, we will not further consider estimation of this model.

\subsubsection{Multiplicative Causal Effect  and  Confounding Bias on the Odds Ratio Scale }

If the causal effect $\tilde{\gamma}$ is defined on the multiplicative scale  and the confounding bias $\tilde{\alpha}$ is defined on the odds ratio scale as follows 
\begin{eqnarray*}
\tilde{\gamma} &=& \frac{E(Y\mid A=1)}{E(Y_0\mid A=1)} \in \mathbb{R}^+,\\
\tilde{\alpha} &=& \frac{E(Y_0\mid A=1)(1-E(Y_0\mid A=0))}{E(Y_0\mid A=0)(1-E(Y_0\mid A=1))}  \in \mathbb{R}^+.
\end{eqnarray*}

\begin{figure}[ht]
	\centering
	\includegraphics[scale=0.8]{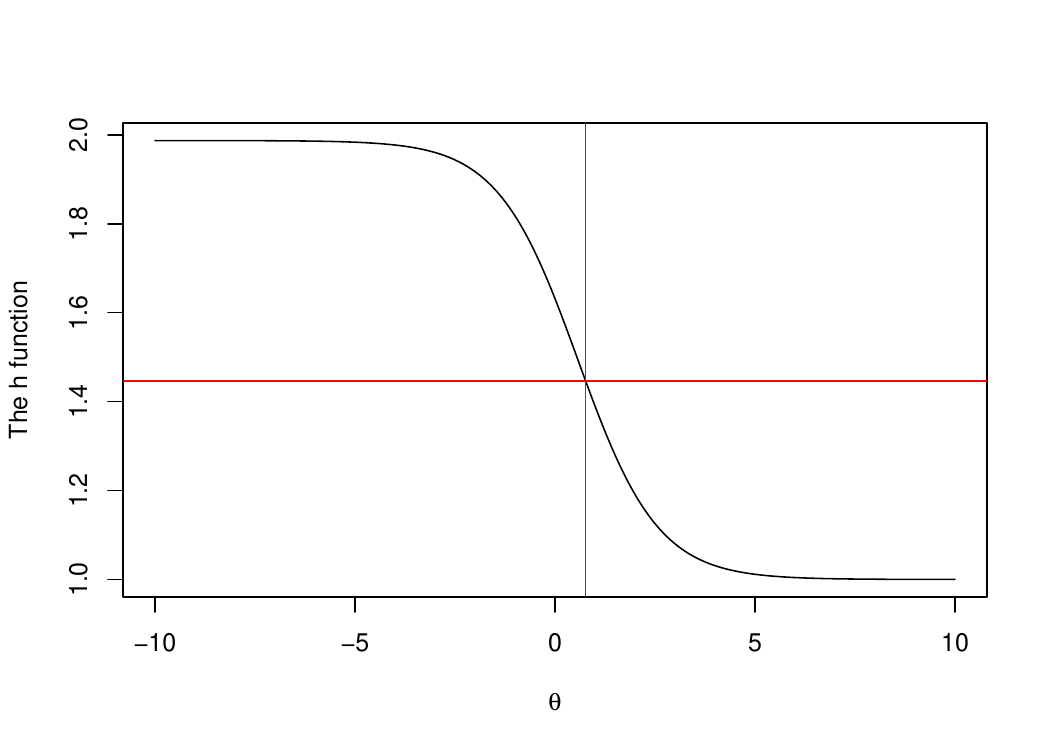}
	\caption{A graphical illustration of the $h$ function. The horizontal axis represents the bias parameter $\theta$  and the vertical axis represents the $h$ function. $E(Y|A=0,Z=1)=0.36$, $E(Y|A=0,Z=0)=0.22$. The red horizontal line represents $E(Y|A=1,Z=1)/E(Y|A=1,Z=0)=1.446$, the red vertical  lines represent the root $0.767$. }
	\label{fig:RR-OR-roots}
\end{figure}

With a possibly invalid instrument $Z$, we again stratify the population to obtain the conditional causal effect $\tilde{\gamma}(z)$ and conditional confounding bias $\tilde{\alpha}(z)$ as follows 
\begin{eqnarray*}
\tilde{\gamma}(z) &=& \frac{E(Y\mid A=1,z)}{E(Y_0\mid A=1,z)} \in \mathbb{R}^+,\\
\tilde{\alpha}(z) & =& \frac{E(Y_0\mid A=1,Z=z)(1-E(Y_0\mid A=0,Z=z))}{E(Y_0\mid A=0,Z=z)(1-E(Y_0\mid A=1,Z=z))}.
\end{eqnarray*}
Assume that $\tilde{\gamma}(z)$ and $\tilde{\alpha}(z)$ do not vary with $z$, then 
we can represent $E(Y_0\mid A=1, Z=z)$ as follows
\[
E(Y_0\mid A=1,Z=z ) = \frac{\tilde{\alpha} E(Y_0\mid A = 0,Z=z)}{\tilde{\alpha} E(Y_0\mid A = 0,Z=z) + 1- E(Y_0\mid A = 0,Z=z) }.
\]
Denote $\tilde{\alpha} = \exp(\theta)$, then we have 
\[
E(Y_0|A=1,Z=z;\theta) =  \frac{\exp(\theta) E(Y\mid A = 0,z)}{\exp(\theta) E(Y \mid A = 0,z) + 1- E(Y\mid A = 0,z) }.
\]
Under the assumption that the causal effect $\tilde{\gamma}(z)$  is constant across the level of $Z$, then 
\begin{equation}
\tilde{\gamma} =  \frac{E(Y_1|A=1,Z=1)}{E(Y_0|A=1,Z=1;\theta)}   =   \frac{E(Y_1|A=1,Z=0)}{E(Y_0|A=1,Z=0;\theta)} .
   \label{eqn:homo-causal}
\end{equation}
Re-arrange the terms in equation (\ref{eqn:homo-causal}), we have the following equation
\begin{equation}
 \frac{E(Y|A=1,Z=1)}{E(Y|A=1,Z=0)}   =   \frac{E(Y_0|A=1,Z=1;\theta)}{E(Y_0|A=1,Z=0;\theta)} .
   \label{eqn:multi-OR-bias-eqn}
\end{equation}
Hence, the parameter $\theta$ can be identified if and only if there is one and only one solution for $\theta$ in equation (\ref{eqn:multi-OR-bias-eqn}). Denote  the right-hand side of equation (\ref{eqn:multi-OR-bias-eqn})  as $h(\theta)$ which is monotone in $\theta$ as shown in  Figure \ref{fig:RR-OR-roots}. The left-hand side of equation (\ref{eqn:multi-OR-bias-eqn}) is a positive number, denoted as $c \in \mathbb{R}^+$, then for identification of $\theta$ we need to show there exist one and only one solution to the following nonlinear equation 
\begin{equation}
    h(\theta) = c.
\end{equation}
Since the $h(\theta)$ function is monotone, then $\theta = h^{-1}(c)$. 
Hence, both the confounding bias parameter $\theta$ on the odds ratio scale and the causal effect on the multiplicative scale $\tilde{\gamma}$ can be  identified.

\section{Conditional Independence Justification for the No-Interaction Assumption}

 Let $\mathcal{R}$ denote the response type of a person, such that $\mathcal{R} =i$ denotes that a given person is immune to developing the outcome $Y$ without treatment, regardless of the value the quasi-instrumental variable (QIV) is set to, that is, $I\{Y_{a=0,z=1}=Y_{a=0,z=0}=0\}=1$. Furthermore, let $\mathcal{R}=d$ if $I\{Y_{a=0,z=1}=Y_{a=0,z=0}=1\}=1$ for a person who is doomed to experience the outcome without treatment, irrespective of the QIV; $\mathcal{R}=h$ if $I\{Y_{a=0,z=1}=1,Y_{a=0,z=0}=0\}=1$ for a person who is harmed by the QIV, in that they would experience the outcome without treatment only if the QIV were set at 1, but not otherwise; and $\mathcal{R}=p$ if $I\{Y_{a=0,z=1}=0,Y_{a=0,z=0}=1\}=1$ for a person that would be protected from experiencing the outcome without treatment only if the QIV were set at 1, but not otherwise. Suppose next that confounding of the causal effect of $A$ on $Y$ is predicated exclusively on the person's disease immunity status, in the sense that any hidden cause $U$ of $A$ that is predictive of disease outcome $Y$, and is therefore a confounder, does so only to the extent that it is associated with the person's immunity status, i.e. it determines whether or not $\mathcal{R} =i$; however, among the subset of disease susceptible persons, i.e. with $\mathcal{R} \neq i$, $U$ is equally distributed across response types, that is:
\[
\mathcal{R}\amalg U|\mathcal{R}\neq i.
\]%
As a consequence, while $U$ confounds the effect of $A$ and $Y$, in principle, one could successfully account for such confounding by restricting the analysis to the subset of susceptible individuals. This is of course not a feasible strategy for confounding adjustment given that response types are not observed; nevertheless, the conditional independence implies that $P \left( Y_{a=0}=1|U=u,Z=1\right)/P \left( Y_{a=0}=1|U=u,Z=0\right)$ is equal to 
\[
\frac{P \left( Y_{a=0,z=1}=1|U=u\right)}{P \left( Y_{a=0,z=0}=1|U=u\right)}=
\frac{P \left( \mathcal{R}\in \{d,h\} |U=u\right)}{P \left( \mathcal{R}\in \{d,p\} |U=u\right)}=\frac{P \left( \mathcal{R}\in \{d,h\} |U=u,\mathcal{R} \neq i\right)}{P \left( \mathcal{R}\in \{d,p\} |U=u, \mathcal{R} \neq i\right)}
\]
which by the conditional independence condition, is equal to \[\frac{P \left( \mathcal{R}\in \{d,h\} |\mathcal{R} \neq i\right)}{P \left( \mathcal{R}\in \{d,p\} | \mathcal{R} \neq i\right)}
\]
and therefore, does not depend on $U$; and therefore the conditional independence condition $\mathcal{R}\amalg U|\mathcal{R}\neq i$ implies $$\beta _{u,z}\left( U,Z\right)=\beta _{u}\left( U\right)\beta _{z}\left(Z\right).$$

\section{Lack of Identification With Parallel Trends on the Odds Ratio Scale}

\citet{liu2022mendelian} considered the odds ratio confounding function:
\[
 \tilde{\alpha}(z,x)= \frac{P(Y_{a=0} =1\mid A=1,Z=z,X=x)P(Y_{a=0}=0\mid A=0,Z=z,X=x)}{P(Y_{a=0}=1\mid A=0,Z=z,X=x)P(Y_{a=0}=0\mid A=1,Z=z,X=x)},
\]
and assumed that it does not vary across levels of $Z$, i.e., $\tilde{\alpha}(z,x) = \tilde{\alpha}(x)$.  To simplify the exposition,  we omit $X$ in this section without loss of generality. Denote $\tilde{\alpha} = \exp(\theta)$ where $\theta \in \mathbb{R}$. The unobserved expectation $E(Y_0\mid A=1,Z=z;\tilde{\alpha})$ can be written as follows \citep{Liu2020Sinica}
\[
 E(Y_{a=0}\mid A=1,Z=z;\theta) = \frac{\exp(\theta) E(Y\mid A = 0,Z=z)}{ \exp(\theta) E(Y \mid A = 0,Z=z) + 1- E(Y\mid A = 0,Z=z) }.
 \]
Under the no current treatment value interaction assumption, we have the following equation 
\begin{equation}
E(Y_{a=1} - Y_{a=0}\mid A=1, Z=1) = E(Y_{a=1} - Y_{a=0}\mid A=1,Z=0).
\label{eqn:assumptionB1}
\end{equation}
After rearranging the terms in equation (\ref{eqn:assumptionB1}), we obtain that  
\begin{equation}
 E(Y_{a=1}\mid A=1,Z=1) - E(Y_{a=1}\mid A=1,Z=0) = g(\theta),
 \label{eqn:g.theta.eqn}
\end{equation}
where the function   $g(\theta)$ is 
\[
g(\theta) = E(Y_{a=0}\mid A=1,Z=1;\theta) -E(Y_{a=0}\mid A=1,Z=0;\theta). 
\]
The left-hand side of equation (\ref{eqn:g.theta.eqn}) is  the observed  mean difference of the outcome among the treated, comparing the $Z=1$  versus the $Z=0$ sub-populations; and the right-hand side of equation  (\ref{eqn:g.theta.eqn}) is not directly observable. Point identification is achieved if there exists one and only one solution for $\theta$ in  equation (\ref{eqn:g.theta.eqn}). 

 However, as we show below, there is in fact no guarantee of a unique solution for $\theta$ in  equation (\ref{eqn:g.theta.eqn}). 
This is because  for any given observed value of the left-hand side of equation (\ref{eqn:g.theta.eqn}), the right-hand side function $g(\theta)$ is not a monotone function in $\theta$ as shown in Figure \ref{fig:ORlinksroots}. As a result, it is in fact possible that there exists two distinct values of $\theta$: $\theta_1 \neq \theta_2$, such that 
\begin{eqnarray*}
 E(Y_{a=1}\mid A=1,Z=1) - E(Y_{a=1}\mid A=1,Z=0) &=& g(\theta_1),\\
  E(Y_{a=1}\mid A=1,Z=1) - E(Y_{a=1}\mid A=1,Z=0) &=& g(\theta_2),
\end{eqnarray*}
which would contradict identification.

\begin{example}
We provide a numerical illustration to demonstrate the lack of identification. For notational simplicity, let \( p_{aa^*}(z) = P(Y_a = 1 \mid A = a^*, Z = z) \). Suppose we observe the following four probabilities: \( p_{00}(z=0) = 0.221 \), \( p_{00}(z=1) = 0.360 \), \( p_{11}(z=0) = 0.337 \), and \( p_{11}(z=1) = 0.487 \). The unobserved probabilities \( p_{01}(z; \theta) \) depend on the bias parameter \( \theta \).

We consider the equation:
\[
p_{11}(z=1) - p_{11}(z=0) = 0.150 = g(\theta),
\]
which admits two solutions for \( \theta \): \( 0.200 \) and \( 1.636 \). This two-root case is illustrated in Figure~\ref{fig:ORlinksroots}.

\begin{figure}[ht]
	\centering
	\includegraphics[scale=0.6]{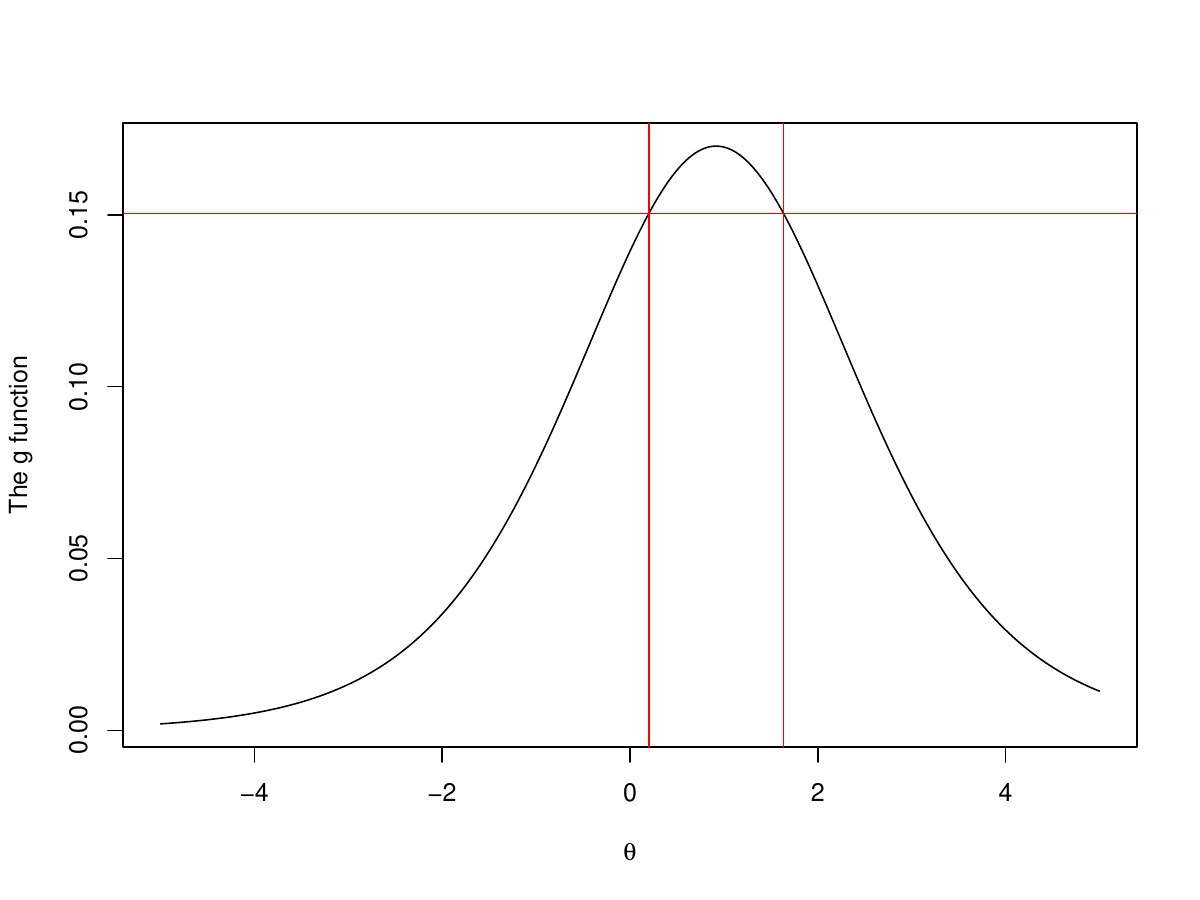}
	\caption{Lack of identification when the confounding bias is defined on the odds ratio scale. The horizontal axis represents the bias parameter \( \theta \), and the vertical axis represents the function \( g(\cdot) \). The red horizontal line corresponds to \( p_{11}(z=1) - p_{11}(z=0) = 0.150 \). The red vertical lines indicate the two roots: \( \theta = 0.200 \) and \( \theta = 1.636 \).}
	\label{fig:ORlinksroots}
\end{figure}

This example illustrates that two distinct parameter configurations can yield the same observed data for \( p_{00}(z) \) and \( p_{11}(z) \), for both \( z = 0 \) and \( z = 1 \). Because \( p_{01}(z) \) is unobserved, it may differ across the two parameter sets, thereby precluding identification. Table~\ref{tab:probsmaintext} summarizes the probability values under both setups. While the observed probabilities are identical, the unobserved \( p_{01}(z) \) values vary, demonstrating the non-identifiability.

\begin{table}[!htbp]
	\centering
	\caption{Probabilities \( p_{00}(z), p_{01}(z), p_{11}(z) \) under two distinct parameter setups. The observed probabilities are numerically identical, but the unobserved probabilities \( p_{01}(z) \) differ across the two configurations.}
	\begin{tabular}{*6c}
		\toprule
		Setup 1 &  \multicolumn{5}{c}{\( \beta = 0.080, \theta = 0.200 \)}  \\
		\midrule
		\( p_{00}(z=0) \) & \( p_{00}(z=1) \) & \( p_{01}(z=0) \) & \( p_{01}(z=1) \) & \( p_{11}(z=0) \) & \( p_{11}(z=1) \) \\
		0.221 & 0.360 & \textit{0.257} & \textit{0.407} & 0.337 & 0.487 \\
		\midrule
		Setup 2 &  \multicolumn{5}{c}{\( \beta = -0.261, \theta = 1.636 \)}  \\
		\midrule
		\( p_{00}(z=0) \) & \( p_{00}(z=1) \) & \( p_{01}(z=0) \) & \( p_{01}(z=1) \) & \( p_{11}(z=0) \) & \( p_{11}(z=1) \) \\
		0.221 & 0.360 & \textit{0.593} & \textit{0.743} & 0.337 & 0.487 \\
		\bottomrule
	\end{tabular}
	\label{tab:probsmaintext}
\end{table}
\end{example}

\section{Technical Proofs}

\subsection{Proof of Theorem 1}
\begin{proof}
First, we have the following  equations
	\begin{eqnarray*}
P(Y=1|A=1,Z=0,X=x) &=&\gamma(x) + \alpha(x) P(Y=1|A=0,Z=0,X=x) ,  \\
P(Y=1|A=1,Z=1,X=x) &=&\gamma(x) + \alpha(x) P(Y=1|A=0,Z=1,X=x).
\end{eqnarray*}
Solving these two equations directly gives the following closed-form solutions
	\begin{eqnarray}
\alpha(x) &=&\frac{P(Y=1|A=1,Z=1,X=x)  -P(Y=1|A=1,Z=0,X=x)} {P(Y=1|A=0,Z=1,X=x)-P(Y=1|A=0,Z=0,X=x)}   \\
\gamma(x)  &=&  P(Y=1|A=1,Z=z,X=x) - \alpha P(Y=1|A=0,Z=z,X=x), z=0,1.  
\end{eqnarray}
This completes the proof. 
\end{proof}

\subsection{Proof of Theorem 2}

\begin{proof}
By definition, suppressing
	dependence on $z,x$,  we have that 
	\begin{eqnarray*}
		\log \mathrm{GOP}\ &=&\log p_{11}-\log \left(1- p_{11}\right) +\log
		p_{01}-\log \left( 1-p_{01}\right) \\
		&&+\log p_{00}-\log \left( 1-p_{00}\right)
	\end{eqnarray*}%
	Now for any $p_{00}\in \left( 0,1\right) $ such that 
	\begin{eqnarray*}
		0 &\leq &p_{11}\leq 1 \\
		0 &\leq &p_{01}\leq 1
	\end{eqnarray*}%
	we have that 
	\begin{eqnarray*}
		\frac{-\gamma }{\alpha } &\leq &p_{00}\leq \frac{1-\gamma }{\alpha } \\
		0 &\leq &p_{00}\leq \frac{1\ }{\alpha },
	\end{eqnarray*}%
	For the model to be non-trivial,we need $\frac{-\gamma }{\alpha }=\left( 1-%
	\frac{p_{11}}{p_{01}}\right) p_{00}\leq 1$ $\ $and $\frac{1-\gamma }{\alpha }%
	>0$ both of which always hold and equivalently 
	\[
	\max \left( 0,\frac{-\gamma }{\alpha }\right) \leq p_{00}\leq \min \left( 
	\frac{1-\gamma }{\alpha },\frac{1\ }{\alpha },1\right) 
	\]
	For any such $p_{00},$let 
	\begin{eqnarray*}
		g\left( p_{00}\right) &=&\log p_{11}-\log \left( 1-p_{11}\right) +\log
		p_{01}-\log \left( 1-p_{01}\right) \\
		&&+\log p_{00}-\log \left( 1-p_{00}\right) -\log \mathrm{GOP} \\
		&=&\log \left( p_{00}\alpha +\gamma \right) -\log \left( 1-p_{00}\alpha
		-\gamma \right) \\
		&&+\log \left( p_{00}\alpha \right) -\log \left( 1-p_{00}\alpha \right) \\
		&&+\log p_{00}-\log \left( 1-p_{00}\right) -\log \mathrm{GOP}
	\end{eqnarray*}%
	therefore%
	\begin{eqnarray*}
		\ \frac{\partial g\left( p_{00}\right) }{\partial p_{00}} &=&\frac{\alpha }{%
			p_{00}\alpha +\gamma }+\frac{\alpha }{1-\left( p_{00}\alpha +\gamma \right) }
		\\
		&&+\frac{2}{p_{00}}+\frac{\alpha }{1-\left( p_{00}\alpha \ \right) }+\frac{1%
		}{1-p_{00}} \\
		&\geq &0
	\end{eqnarray*}
	
	Next note that as $p_{00}\rightarrow \max \left( 0,-\frac{-\gamma }{\alpha }%
	\right) $ from the right, we have that $g\left( p_{00}\right) \rightarrow
	-\infty ,$ likewise, if $p_{00}\rightarrow \min \left( \frac{1-\gamma }{%
		\alpha },\frac{1\ }{\alpha },1\right) $ from the left$,$ we have that $%
	p_{00}\rightarrow \infty .$ Therefore given that $\frac{\partial g\left(
		p_{00}\right) }{\partial p_{00}}\geq 0$ for all $\max \left( 0,\frac{-\gamma 
	}{\alpha }\right) \leq p_{00}\leq \min \left( \frac{1-\gamma }{\alpha },%
	\frac{1\ }{\alpha },1\right) ,$ we can conclude that there is only one root
	for $g\left( p_{00}\right) =0$ crosses zero only ones in this interval.
	Since the domain of $\left( \gamma ,\alpha ,\mathrm{GOP}\right) \in \left(
	-1,1\right) \times \mathbb{R}^{+}\times \mathbb{R}^{+}$ is the cartesian
	product of the domains of $\gamma ,\alpha $ and $\mathrm{GOP}$\textrm{, }the
	models are variation independent. 
\end{proof}

\subsection{Closed-Form Solution for \texorpdfstring{$p_{00}(z,x)$}{p00(z,x)} via Cardano's Method}

We now derive a closed-form expression for the baseline risk \(p_{00}(z,x)\), 
which appears in the additive–multiplicative model for binary outcomes introduced in the main text. 
The derivation follows the classical method of Cardano \citep{abramowitz1965handbook} for solving cubic equations. 

Specifically, \(p_{00}(z,x)\) satisfies the cubic polynomial
\[
b_1(z,x)\, p_{00}^3(z,x) + b_2(z,x)\, p_{00}^2(z,x) + b_3(z,x)\, p_{00}(z,x) + b_4(z,x) = 0.
\]

To simplify, we apply the standard change of variables
\[
p_{00}(z,x) = t - \frac{b_2(z,x)}{3b_1(z,x)},
\]
which reduces the cubic to its depressed form
\[
t^3 + \xi t + \zeta = 0,
\]
with coefficients
\[
\xi = \frac{3b_1(z,x)b_3(z,x) - b_2(z,x)^2}{3b_1(z,x)^2}, 
\qquad 
\zeta = \frac{2b_2(z,x)^3 - 9b_1(z,x)b_2(z,x)b_3(z,x) + 27b_1(z,x)^2b_4(z,x)}{27b_1(z,x)^3}.
\]

The discriminant of this cubic is
\[
\Delta = \left( \frac{\zeta}{2} \right)^2 + \left( \frac{\xi}{3} \right)^3.
\]

\begin{itemize}
\item \textbf{Case 1: \(\Delta \geq 0\)}.  
There is a unique real root, given by Cardano's formula:
\[
t = \sqrt[3]{-\frac{\zeta}{2} + \sqrt{\Delta}} 
  + \sqrt[3]{-\frac{\zeta}{2} - \sqrt{\Delta}},
\]
and thus
\[
p_{00}(z,x) = t - \frac{b_2(z,x)}{3b_1(z,x)}.
\]

\item \textbf{Case 2: \(\Delta < 0\)}.  
There are three distinct real roots, expressed via the trigonometric (casus irreducibilis) solution:
\[
\theta = \cos^{-1}\!\left( \frac{3\zeta}{2\xi} \sqrt{-\frac{3}{\xi}} \right),
\qquad
t_k = 2 \sqrt{-\frac{\xi}{3}} \cos\!\left( \frac{\theta + 2k\pi}{3} \right), \quad k=0,1,2,
\]
\[
p_{00}^{(k)}(z,x) = t_k - \frac{b_2(z,x)}{3b_1(z,x)}.
\]
Among these, the root \(p_{00}^{(k)}(z,x)\) falling within the unit interval \((0,1)\) is selected, ensuring compatibility with the binary outcome.
\end{itemize}

This closed-form derivation confirms that the model-implied probabilities \(p_{00}(z,x)\), 
\(p_{01}(z,x)\), and \(p_{11}(z,x)\) all lie within \((0,1)\), thereby preserving variation independence 
and ensuring statistical coherence of the proposed parameterization.

\subsection{Proof of Theorem 3}

Consider a parametric path $\{F_t: t\}$ for the joint distribution of $O=(Y,A,Z,X)$ which is equal to the true distribution $F_0$ when $t=0$ with corresponding score $S(O)=\partial \ln (d F_t)/\partial t$, where derivatives are evaluated at $t=0$ unless otherwise indicated. Following \citet{newey1990semiparametric} and \citet{bickel1993efficient}, we derive the efficient influence function (EIF) in the nonparametric observed data model as the pathwise derivative $\psi(O;\gamma)$ satisfying $$\partial \gamma_t/\partial t = E\{\psi(O;\gamma)S(O)\}.$$
To simplify the proof, we introduce the following lemma.

\begin{lemma}
\label{lem:1}
 Consider a product parameter $\tau=\tau_1 \tau_2$.  The EIF for $\tau$ is $\tau_1\psi_2(O)+\tau_2\psi_1(O)$ if and only if $\psi_1(O)$ and $\psi_2(O)$ are the EIFs for $\tau_1$ and $\tau_2$, respectively.   
\end{lemma}

\begin{proof}[Proof of Lemma \ref{lem:1}]
By the product rule,
\begin{equation*}
    \begin{aligned}
 \frac{\partial \tau_t}{\partial t}&=\tau_1 \frac{\partial \tau_{2,t}}{\partial t}+\tau_2 \frac{\partial \tau_{1,t}}{\partial t}\\
&=\tau_1 E\{\psi_2(O)S(O)\}+\tau_2 E\{\psi_1(O)S(O)\}\\
&=E[\{\tau_1\psi_2(O)+\tau_2\psi_1(O)\}S(O)].       
    \end{aligned}
\end{equation*}
\end{proof}

\begin{proof}[Proof of Theorem 3]
By Theorem 1, 
\[
0=E\left\{ p\left( A=1|X,Z\right) \frac{\left\{ YA-Y(1-A)\alpha \left(
X\right) \right\} }{p\left( A|Z,X\right) }-\gamma A\right\}. 
\]%
Therefore,  
\begin{eqnarray*}
&&\nabla _{t}E_{t}\left\{ p_{t}\left( A=1|X,Z\right) \frac{\left\{
YA-Y(1-A)\alpha _{t}\left( X\right) \right\} }{p_{t}\left( A|Z,X\right) }%
-\gamma _{t}A\right\}  \\
&=&E\left\{ S(O)\left[p\left( A=1|X,Z\right) \frac{\left\{ YA-Y(1-A)\alpha \left(
X\right) \right\} }{p\left( A|Z,X\right) }-\gamma A\right]\right\}  \\
&&+E\left\{ \left( \nabla _{t}p_{t}\left( A=1|X,Z\right) E\left[ \frac{%
\left\{ YA-Y(1-A)\alpha \left( X\right) \right\} }{p\left( A|Z,X\right) }\biggr\rvert X,Z%
\right] \right) \right\}  \\
&&+E\left\{ \left[p\left( A=1|X,Z\right) \nabla _{t}\frac{\left\{ E\left(
Y|A,X,Z\right) A-E\left( Y|A,X,Z\right) (1-A)\alpha \left( X\right) \right\} 
}{p_{t}\left( A|Z,X\right) }\right]\right\}  \\
&&-E\left[ E\left\{ p\left( A=1|X,Z\right) \left[ E\left( Y|A=0,X,Z\right) %
\right] |X\right\} \nabla _{t}\alpha _{t}\left( X\right) \right]  \\
&&-\nabla _{t}\gamma _{t}p \left( A=1\right). 
\end{eqnarray*}
Note that 
\begin{eqnarray*}
&&E\left\{ \left( \nabla _{t}p_{t}\left( A=1|X,Z\right) E\left[ \frac{%
\left\{ YA-Y(1-A)\alpha \left( X\right) \right\} }{p\left( A|Z,X\right) }\biggr\rvert X,Z%
\right] \right) \right\}  \\
&=&E\left( \nabla _{t}p_{t}\left( A=1|X,Z\right) \gamma \left( X\right)
\right)  \\
&=&E\left\{ \gamma \left( X\right) \left( A-p \left( A=1|X,Z\right)
\right) S(A|X,Z)\right\}, 
\end{eqnarray*}%
and%
\begin{eqnarray*}
&&E\left\{ \left[p\left( A=1|X,Z\right) \nabla _{t}\frac{\left\{ E\left(
Y|A,X,Z\right) A-E\left( Y|A,X,Z\right) (1-A)\alpha \left( X\right) \right\} 
}{p_{t}\left( A|Z,X\right) }\right]\right\}  \\
&=&-E\left\{ \left[\frac{\nabla _{t}p_{t}\left( A|Z,X\right) }{p\left(
A|Z,X\right) }p\left( A=1|X,Z\right) \frac{\left\{ E\left( Y|A,X,Z\right)
A-E\left( Y|A,X,Z\right) (1-A)\alpha \left( X\right) \right\} }{p\left(
A|Z,X\right) }\right]\right\}  \\
&=&-E\left\{ S\left( A|Z,X\right) \frac{p\left( A=1|X,Z\right) \left\{
E\left( Y|A,X,Z\right) A-E\left( Y|A,X,Z\right) (1-A)\alpha \left( X\right)
\right\} }{p\left( A|Z,X\right) }\right\}  \\
&=&-E\left\{ S\left( A|Z,X\right) \left[ 
\begin{array}{c}
\frac{p\left( A=1|X,Z\right) \left\{ E\left( Y|A,X,Z\right) A-E\left(
Y|A,X,Z\right) (1-A)\alpha \left( X\right) \right\} }{p\left( A|Z,X\right) }
\\ 
-\gamma \left( X\right) p\left( A=1|X,Z\right) 
\end{array}%
\right] \right\}. 
\end{eqnarray*}%
Finally consider the last term  
\[
E\left[ E\left\{ p\left( A=1|X,Z\right) \left[ E\left( Y|A=0,X,Z\right) %
\right] |X\right\} \nabla _{t}\alpha _{t}\left( X\right) \right] 
\]%
where $\alpha _{t}\left( X\right) $ satisfies%
\begin{eqnarray*}
0 &=&E_{t}\left( Y|A=1,Z=1,X\right) -E_{t}\left( Y|A=1,Z=0,X\right)  \\
&&-\alpha _{t}\left( X\right) \left[ E_{t}\left( Y|A=0,Z=1,X\right)
-E_{t}\left( Y|A=0,Z=0,X\right) \right]. 
\end{eqnarray*}%
Therefore 
\begin{eqnarray*}
0 &=&\nabla _{t}E_{t}\left( Y|A=1,Z=1,X\right) -\nabla _{t}E_{t}\left(
Y|A=1,Z=0,X\right)  \\
&&-\nabla _{t}\alpha _{t}\left( X\right) \left[ E\left( Y|A=0,Z=1,X\right)
-E\left( Y|A=0,Z=0,X\right) \right]  \\
&&-\alpha \left( X\right) \left[ \nabla _{t}E_{t}\left( Y|A=0,Z=1,X\right)
-\nabla _{t}E_{t}\left( Y|A=0,Z=0,X\right) \right]  \\
&=&E\left\{ \left\{ Y-E\left( Y|A,Z,X\right) \right\} \frac{\left( -1\right)
^{A+Z}}{f\left( A,Z|X\right) }\alpha \left( X\right)
^{1-A}S(Y,A,Z|X)|X\right\}  \\
&=&E\left\{ \left\{ Y-E\left( Y|A,Z,X\right) \right\} \frac{\left( -1\right)
^{A+Z}}{f\left( A,Z|X\right) }\alpha \left( X\right)
^{1-A}S(Y,A,Z|X)|X\right\}  \\
&&-\nabla _{t}\alpha _{t}\left( X\right) \left[ E\left( Y|A=0,Z=1,X\right)
-E\left( Y|A=0,Z=0,X\right) \right], 
\end{eqnarray*}%
which implies that%
\begin{eqnarray*}
&&\nabla _{t}\alpha _{t}\left( X\right)  \\
&=&E\left\{ \frac{\left\{ Y-E\left( Y|A,Z,X\right) \right\} }{\left[ E\left(
Y|A=0,Z=1,X\right) -E\left( Y|A=0,Z=0,X\right) \right] }\frac{\left(
-1\right) ^{A+Z}}{f\left( A,Z|X\right) }\alpha \left( X\right)
^{1-A}S(Y,A,Z|X)\biggr\rvert X\right\}. 
\end{eqnarray*}%
Therefore, since   
\begin{eqnarray*}
0 &=&\nabla _{t}E_{t}\left\{ p_{t}\left( A=1|X,Z\right) \frac{\left\{
YA-Y(1-A)\alpha _{t}\left( X\right) \right\} }{p_{t}\left( A|Z,X\right) }%
-\gamma _{t}A\right\}  \\
&=&E\left\{ S(O)[p\left( A=1|X,Z\right) \frac{\left\{ YA-Y(1-A)\alpha \left(
X\right) \right\} }{p\left( A|Z,X\right) }-\gamma A]\right\}  \\
&&+E\left\{ \left( \nabla _{t}p_{t}\left( A=1|X,Z\right) E\left[ \frac{%
\left\{ YA-Y(1-A)\alpha \left( X\right) \right\} }{p\left( A|Z,X\right) }|X,Z%
\right] \right) \right\}  \\
&&+E\left\{ [p\left( A=1|X,Z\right) \nabla _{t}\frac{\left\{ E\left(
Y|A,X,Z\right) A-E\left( Y|A,X,Z\right) (1-A)\alpha \left( X\right) \right\} 
}{p_{t}\left( A|Z,X\right) }]\right\}  \\
&&-E\left[ E\left\{ p\left( A=1|X,Z\right) \left[ E\left( Y|A=0,X,Z\right) %
\right] |X\right\} \nabla _{t}\alpha _{t}\left( X\right) \right]  \\
&&-\nabla _{t}\gamma _{t}p \left( A=1\right), 
\end{eqnarray*}
and 
\bigskip 
\begin{eqnarray*}
&&\nabla _{t}\gamma _{t}p \left( A=1\right)  \\
&=&E\left\{ S(O)[p\left( A=1|X,Z\right) \frac{\left\{ YA-Y(1-A)\alpha \left(
X\right) \right\} }{p\left( A|Z,X\right) }-\gamma A]\right\}  \\
&&+E\left\{ S(O)\gamma \left( X\right) \left( A-p \left( A=1|X,Z\right)
\right) \right\}  \\
&&-E\left\{ S\left( O\right) \left[ \frac{p\left( A=1|X,Z\right) \left\{
E\left( Y|A,X,Z\right) A-E\left( Y|A,X,Z\right) (1-A)\alpha \left( X\right)
\right\} }{p\left( A|Z,X\right) }-\gamma \left( X\right) p\left(
A=1|X,Z\right) \right] \right\}  \\
&&-E\left[ S(O)E\left\{ 
\begin{array}{c}
\frac{E\left\{ p\left( A=1|X,Z\right) \left[ E\left( Y|A=0,X,Z\right) \right]
|X\right\} \left\{ Y-E\left( Y|A,Z,X\right) \right\} }{\left[ E\left(
Y|A=0,Z=1,X\right) -E\left( Y|A=0,Z=0,X\right) \right] } \\ 
\times \frac{\left( -1\right) ^{A+Z}}{f\left( A,Z|X\right) }\alpha \left(
X\right) ^{1-A}S(Y,A,Z|X)|X%
\end{array}%
\right\} \right], 
\end{eqnarray*}
the EIF\ of $\gamma p \left( A=1\right) $ is given by 
\begin{eqnarray*}
&&p\left( A=1|X,Z\right) \frac{\left\{ YA-Y(1-A)\alpha \left( X\right)
\right\} }{p\left( A|Z,X\right) }-\gamma A \\
&&+\gamma \left( X\right) \left( A-p \left( A=1|X,Z\right) \right)  \\
&&-\left[ \frac{p\left( A=1|X,Z\right) \left\{ E\left( Y|A,X,Z\right)
A-E\left( Y|A,X,Z\right) (1-A)\alpha \left( X\right) \right\} }{p\left(
A|Z,X\right) }-\gamma \left( X\right) p\left( A=1|X,Z\right) \right]  \\
&&-E\left\{ p\left( A=1|X,Z\right) \left[ E\left( Y|A=0,X,Z\right) \right]
|X\right\}  \\
&&\times \left\{ \frac{\left\{ Y-E\left( Y|A,Z,X\right) \right\} }{\left[
E\left( Y|A=0,Z=1,X\right) -E\left( Y|A=0,Z=0,X\right) \right] }\frac{\left(
-1\right) ^{A+Z}}{f\left( A,Z|X\right) }\alpha \left( X\right)
^{1-A}\right\}. 
\end{eqnarray*}
The EIF for $p(A=1)$ in the nonparametric observed data model is $A-p(A=1)$. Then by Lemma \ref{lem:1}, the efficient influence function  $\psi(O;\gamma)$ for estimating $\gamma$ is given by
\begin{eqnarray*}
&& \frac{1}{p(A=1)}\Biggr(p\left( A=1|X,Z\right) \frac{\left\{ YA-Y(1-A)\alpha \left( X\right)
\right\} }{p\left( A|Z,X\right) } +\gamma \left( X\right) \left( A-p \left( A=1|X,Z\right) \right)  \\
&&-\left[ \frac{p\left( A=1|X,Z\right) \left\{ E\left( Y|A,X,Z\right)
A-E\left( Y|A,X,Z\right) (1-A)\alpha \left( X\right) \right\} }{p\left(
A|Z,X\right) }-\gamma \left( X\right) p\left( A=1|X,Z\right) \right]  \\
&&-E\left\{ p\left( A=1|X,Z\right) \left[ E\left( Y|A=0,X,Z\right) \right]
|X\right\}  \\
&&\times \left\{ \frac{\left\{ Y-E\left( Y|A,Z,X\right) \right\} }{\left[
E\left( Y|A=0,Z=1,X\right) -E\left( Y|A=0,Z=0,X\right) \right] }\frac{\left(
-1\right) ^{A+Z}}{p\left( A,Z|X\right) }\alpha \left( X\right)
^{1-A}\right\} \Biggr)-\gamma.
\end{eqnarray*}
\end{proof}

\subsection*{Proof of Theorem 4}
\begin{proof}
In the following, we use the superscript $\ast$ to denote the probability limit of first-step estimator under a possibly misspecified model. Suppose the models for $\alpha \left( X\right) $ and $p\left(
Z,A|X\right) $ are correct, then 
\begin{eqnarray*}
&&E\left\{ p\left( A=1|X,Z\right) \frac{\left\{ YA-Y(1-A)\alpha \left(
X\right) \right\} }{p\left( A|Z,X\right) }-\gamma A\right.  \\
&&+\gamma \left( X\right) \left( A-p \left( A=1|X,Z\right) \right)  \\
&&-\left[ 
\begin{array}{c}
\frac{p\left( A=1|X,Z\right) \left\{ \ \left( \gamma ^{\ast }\left( X\right)
+p_{00}(X,Z)\alpha \left( X\right) \right) A-p_{00}(X,Z)\alpha \left(
X\right) (1-A)\right\} }{p\left( A|Z,X\right) } \\ 
-\gamma ^{\ast }\left( X\right) p\left( A=1|X,Z\right) 
\end{array}%
\right]  \\
&&\left. -E\left\{ p\left( A=1|X,Z\right) \left[ p_{00}(X,Z)\right]
|X\right\} \left\{ 
\begin{array}{c}
\frac{\left\{ Y-\left( \gamma ^{\ast }\left( X\right) A+p_{00}(X,Z)\alpha
\left( X\right) ^{A}\right) \right\} }{\left[ p_{00}(X,Z=1)-p_{00}(X,Z=0)%
\right] } \\ 
\times \frac{\left( -1\right) ^{A+Z}}{p\left( A,Z|X\right) }\alpha \left(
X\right) ^{1-A}%
\end{array}%
\right\} \right\}  \\
&=&E\left\{ -\left[ p\left( A=1|X,Z\right) \left\{ \ \gamma ^{\ast }\left(
X\right) \right\} -\gamma \left( X\right) p\left( A=1|X,Z\right) \right]
\right.  \\
&&\left. -E\left\{ p\left( A=1|X,Z\right) \left[ p_{00}(X,Z)\right]
|X\right\} \left\{ 
\begin{array}{c}
\frac{\left\{ \gamma \left( X\right) A-\left( \gamma ^{\ast }\left( X\right)
A\right) \right\} }{\left[ p_{00}(X,Z=1)-p_{00}(X,Z=0)\right] } \\ 
\times \frac{\left( -1\right) ^{A+Z}}{p\left( A,Z|X\right) }\alpha \left(
X\right) ^{1-A}%
\end{array}%
\right\} \right\}  \\
&=&E\left\{ -\left[ p\left( A=1|X,Z\right) \gamma ^{\ast }\left( X\right)
+\gamma \left( X\right) p\left( A=1|X,Z\right) \right] \right.  \\
&&\left. -E\left\{ p\left( A=1|X,Z\right) \left[ p_{00}(X,Z)\right]
|X\right\} \left\{ \frac{\left\{ \gamma \left( X\right) - \gamma
^{\ast }\left( X\right)\right\} }{\left[ p_{00}(X,Z=1)-p_{00}(X,Z=0)%
\right] }\frac{\left( -1\right) ^{1+Z}}{p\left( Z|X\right) }\right\}
\right\}  \\
&=&E\left\{ -\left[ p\left( A=1|X,Z\right) \gamma ^{\ast }\left( X\right)
+\gamma \left( X\right) p\left( A=1|X,Z\right) \right] \right.  \\
&&\left. -E\left\{ \ \left[ p_{00}(X,Z)\right] |X\right\} \left\{ \frac{%
\left\{ \gamma \left( X\right) p\left( A=1|X,Z\right) - \gamma ^{\ast
}\left( X\right) p\left( A=1|X,Z\right)  \right\} }{\left[
p_{00}(X,Z=1)-p_{00}(X,Z=0)\right] }\frac{\left( -1\right) ^{1+Z}}{p\left(
Z|X\right) }\ \right\} \right\}  \\
&=&E\left\{ -\left[ p\left( A=1|X,Z\right) \gamma ^{\ast }\left( X\right)
+\gamma \left( X\right) p\left( A=1|X,Z\right) \right] \right.  \\
&&\left. -\left\{  \gamma \left( X\right) p\left( A=1|X,Z\right)
-\left( \gamma ^{\ast }\left( X\right) p\left( A=1|X,Z\right) \right)
 \ \right\} \right\}  \\
&=&0.
\end{eqnarray*}
Now suppose that the models for $\gamma \left( X\right) ,p_{00}(X,Z),p(A,Z|X)$ are correct, 
then%
\begin{eqnarray*}
&&E\left\{ p\left( A=1|X,Z\right) \frac{\left\{ YA-Y(1-A)\alpha ^{\ast
}\left( X\right) \right\} }{p\left( A|Z,X\right) }-\gamma A\right.  \\
&&+\gamma \left( X\right) \left( A-p\left( A=1|X,Z\right) \right)  \\
&&-\biggr[ \frac{p\left( A=1|X,Z\right) \left\{ \ \left( \gamma \left(
X\right) +p_{00}(X,Z)\alpha ^{\ast }\left( X\right) \right)
A-p_{00}(X,Z)\alpha ^{\ast }\left( X\right) (1-A)\right\} }{p\left(
A|Z,X\right) }\\
&&-\gamma \left( X\right) p^{\ast }\left( A=1|X,Z\right) \biggr] 
\\
&&\left. -E\left\{ p\left( A=1|X,Z\right) \left[ p_{00}(X,Z)\right]
|X\right\}\times \right.\\ 
&&\left. \left\{ \frac{\left\{ Y-\left( \gamma \left( X\right)
A+p_{00}(X,Z)\alpha ^{\ast }\left( X\right) ^{A}\right) \right\} }{\left[
p_{00}(X,Z=1)-p_{00}(X,Z=0)\right] }\frac{\left( -1\right) ^{A+Z}}{p\left(
A,Z|X\right) }\alpha ^{\ast }\left( X\right) ^{1-A}\right\} \right\}  \\
&=&E\left\{ p\left( A=1|X,Z\right) \left\{ \left( p_{00}(X,Z)\alpha \left(
X\right) \right) -p_{00}(X,Z)\alpha ^{\ast }\left( X\right) \right\} \right. 
\\
&&\left. -E\left\{ p\left( A=1|X,Z\right) \left[ p_{00}(X,Z)\right]
|X\right\} \times \right.\\
&&\left. \left\{ \frac{\left\{ p_{00}(X,Z)\alpha \left( X\right)
^{A}-p_{00}(X,Z)\alpha ^{\ast }\left( X\right) ^{A}\right\} }{\left[
p_{00}(X,Z=1)-p_{00}(X,Z=0)\right] }\frac{\left( -1\right) ^{A+Z}}{p\left(
A|X,Z\right) p\left( Z|X\right) }\alpha ^{\ast }\left( X\right)
^{1-A}\right\} \right\}  \\
&=&E\left\{ p\left( A=1|X,Z\right) \left\{ \left( p_{00}(X,Z)\alpha \left(
X\right) \right) -p_{00}(X,Z)\alpha ^{\ast }\left( X\right) \right\} \right. 
\\
&&-E\left\{ p\left( A=1|X,Z\right) \left[ p_{00}(X,Z)\right] |X\right\}
\left\{ \frac{\left\{ p_{00}(X,Z)\alpha \left( X\right) -p_{00}(X,Z)\alpha
^{\ast }\left( X\right) \right\} }{\left[ p_{00}(X,Z=1)-p_{00}(X,Z=0)\right] 
}\frac{\left( -1\right) ^{1+Z}}{p\left( Z|X\right) }\right\}  \\
&&\left. -E\left\{ p\left( A=1|X,Z\right) \left[ p_{00}(X,Z)\right]
|X\right\} \left\{ \frac{\left\{ p_{00}(X,Z)-p_{00}(X,Z)\right\} }{\left[
p_{00}(X,Z=1)-p_{00}(X,Z=0)\right] }\frac{\left( -1\right) ^{Z}}{p\left(
Z|X\right) }\alpha ^{\ast }\left( X\right) \right\} \right\}  \\
&=&E\left\{ p\left( A=1|X,Z\right) \left\{ \left( p_{00}(X,Z)\alpha \left(
X\right) \right) -p_{00}(X,Z)\alpha ^{\ast }\left( X\right) \right\} \right. 
\\
&&-E\left\{ p\left( A=1|X,Z\right) \left[ p_{00}(X,Z)\right] |X\right\}
\left\{ \left\{ \alpha \left( X\right) -\alpha ^{\ast }\left( X\right)
\right\} \right\}  \\
&=&0
\end{eqnarray*}

Finally, suppose that $p \left( Y=1|A=0,X,Z\right) ,\gamma \left( X\right)
,\alpha \left( X\right) $ are correct, then  
\begin{eqnarray*}
&&E\left\{ p^{\ast }\left( A=1|X,Z\right) \frac{\left\{ YA-Y(1-A)\alpha
\left( X\right) \right\} }{p^{\ast }\left( A|Z,X\right) }-\gamma A\right.  \\
&&+\gamma \left( X\right) \left( A-p^{\ast }\left( A=1|X,Z\right) \right)  \\
&&-\left[ 
\begin{array}{c}
\frac{p^{\ast }\left( A=1|X,Z\right) \left\{ \ \left( \gamma \left( X\right)
+p_{00}(X,Z)\alpha \left( X\right) \right) A-p_{00}(X,Z)\alpha \left(
X\right) (1-A)\right\} }{p^{\ast }\left( A|Z,X\right) } \\ 
-\gamma \left( X\right) p^{\ast }\left( A=1|X,Z\right) 
\end{array}%
\right]  \\
&&\left. -E\left\{ p^{\ast }\left( A=1|X,Z\right) \left[ p_{00}(X,Z)\right]
|X\right\} \left\{ 
\begin{array}{c}
\frac{\left\{ Y-\left( \gamma \left( X\right) A+p_{00}(X,Z)\alpha \left(
X\right) ^{A}\right) \right\} }{\left[ p_{00}(X,Z=1)-p_{00}(X,Z=0)\right] }
\\ 
\times \frac{\left( -1\right) ^{A+Z}}{p\left( A,Z|X\right) }\alpha \left(
X\right) ^{1-A}%
\end{array}%
\right\} \right\}  \\
&=&E\left\{ 
\begin{array}{c}
p\left( A=1|X,Z\right) \left\{ \left( \gamma \left( X\right)
+p_{00}(X,Z)\alpha \left( X\right) \right) \right\}  \\ 
-\frac{p^{\ast }\left( A=1|X,Z\right) p\left( A=0|Z,X\right) }{p^{\ast
}\left( A=0|Z,X\right) }p_{00}(X,Z)\alpha \left( X\right) -\gamma A%
\end{array}%
\right.  \\
&&+\gamma \left( X\right) \left( p\left( A=1|X,Z\right) -p^{\ast }\left(
A=1|X,Z\right) \right)  \\
&&-\left[ 
\begin{array}{c}
\left\{ \ \left( \gamma \left( X\right) +p_{00}(X,Z)\alpha \left( X\right)
\right) p\left( A=1|Z,X\right) \right\}  \\ 
-\frac{p^{\ast }\left( A=1|X,Z\right) p\left( A=0|Z,X\right) }{p^{\ast
}\left( A=0|Z,X\right) }p_{00}(X,Z)\alpha \left( X\right) -\gamma \left(
X\right) p^{\ast }\left( A=1|X,Z\right) 
\end{array}%
\right]  \\
&=&0.
\end{eqnarray*}
\end{proof}

\subsection{Extension to Many Weak QIVs}

In settings where multiple quasi-instruments (QIVs) are available, it is natural to consider combining them to improve estimation precision. In principle, our maximum likelihood estimator (MLE) from Section 3.2 can be extended to accommodate a vector of QIVs, $Z = (Z_1, ..., Z_m)^\top$, by modifying the log-likelihood to incorporate all instruments jointly. The Fisher information matrix from the score function can be used to assess the collective strength of the instruments. Following \citet{liu2022mendelian}, consistency and asymptotic normality of the MLE can be established under the condition that the scaled minimum eigenvalue of the information matrix, $\lambda_{\min}\{nI_1(\phi)\}/k$, diverges as $n \to \infty$.

For the semiparametric estimator described in Section 3.3, a natural extension involves adopting a generalized method of moments (GMM) approach to incorporate multiple moment conditions, each indexed by a distinct QIV or transformation thereof. Although we do not pursue this generalization in the current paper, it offers a promising direction for future work—especially in Mendelian randomization applications involving many weak and possibly invalid SNPs.

Importantly, when the number of QIVs is large relative to sample size, regularization techniques (e.g., penalized likelihood, shrinkage GMM, or machine learning-based moment selection) may be required to ensure statistical and computational tractability. Developing robust identification and estimation theory in such high-dimensional regimes remains a rich and open area for future research.

\section{Additional UK Biobank Data Analysis: 2SLS Results}

We compared two-stage least squares (2SLS) estimates using each SNP individually and all three SNPs jointly. Using all three SNPs produced an estimate of $0.17$ (SE = $0.10$; 95\% CI: $-0.01$, $0.36$; $p=0.07$), but inclusion of the weak SNP (rs11126666) inflated the variance. Single-SNP analyses yielded estimates of $0.01$ (SE = $0.13$; 95\% CI: $-0.24$, $0.26$) for rs3817334, $0.29$ (SE = $0.14$; 95\% CI: $0.01$, $0.58$) for rs3888190, and an unstable estimate of $11.03$ (SE = $17.80$; 95\% CI: $-23.86$, $45.92$) for rs11126666, consistent with weak identification. 

\bigskip

\begin{table}[htbp]
\centering
\caption{Two-stage least squares (2SLS) estimates using each SNP individually and all three SNPs jointly.}
\label{tab:2sls_results}
\resizebox{\columnwidth}{!}{%
\begin{tabular}{l l r r r r r}
\toprule
Instrument   & Model             & Estimate & Std. Error & p-value & 95\% CI (lower) & 95\% CI (upper) \\
\midrule
all\_three   & 2SLS\_all\_three  & 0.173    & 0.0953     & 0.070   & -0.014          & 0.360 \\
rs11126666   & 2SLS\_rs11126666  & 11.029   & 17.8033    & 0.536   & -23.865         & 45.923 \\
rs3817334    & 2SLS\_rs3817334   & 0.0098   & 0.1297     & 0.940   & -0.244          & 0.264 \\
rs3888190    & 2SLS\_rs3888190   & 0.295    & 0.1437     & 0.040   & 0.013           & 0.576 \\
\bottomrule
\end{tabular}
}
\end{table}

\vspace{-0.3in}
\begin{singlespace}
	\bibliographystyle{biom}
\bibliography{QIV}
\end{singlespace}

\end{document}